\documentclass[12pt]{article}
\pdfoutput=1
\usepackage{epsf,amsfonts,amssymb,epsfig,amsmath,mathtools,graphics,slashed}
\usepackage{hyperref,hep,graphicx,subfig}

\newcommand{\nn}{\notag \\}

\makeatletter

\@addtoreset{equation}{section}
\makeatother

\begin{document}

\begin{titlepage}

\vfill

\begin{flushright}
Imperial/TP/2015/JG/04\\
DCPT-15/61
\end{flushright}

\vfill

\begin{center}
   \baselineskip=16pt
   {\Large\bf DC Conductivity of Magnetised\\Holographic Matter}
  \vskip 1.5cm
  \vskip 1.5cm
      Aristomenis Donos$^1$, Jerome P. Gauntlett$^2$, Tom Griffin$^2$ and Luis Melgar$^2$\\
   \vskip .6cm
         \vskip .6cm
      \begin{small}
      \begin{small}
      \textit{$^1$Centre for Particle Theory\\ and Department of Mathematical Sciences\\Durham University, Durham, DH1 3LE, U.K.}\\
          \vskip .6cm
      \textit{$^2$Blackett Laboratory, 
        Imperial College\\London, SW7 2AZ, U.K.}
        \end{small}\\
        \end{small}
\end{center}

\vfill

\begin{center}
\textbf{Abstract}
\end{center}

\begin{quote}
We consider general black hole solutions of Einstein-Maxwell-scalar theory that are holographically 
dual to conformal field theories at finite charge density with non-vanishing magnetic fields 
and local magnetisation currents, which generically
break translation invariance explicitly. We show that the thermoelectric DC conductivity of the field theory
can be
obtained by solving a system of generalised Stokes equations on the black hole horizon.
For various examples, including Q-lattices and one-dimensional lattices, we solve the Stokes equations
explicitly and obtain expressions for the
DC conductivity in terms of the solution at the black hole horizon. 
\end{quote}

\vfill

\end{titlepage}

\setcounter{equation}{0}

\section{Introduction}

In seeking applications of holography to quantum critical systems observed in Nature an important observable to study is the thermoelectric conductivity. At finite temperature the quantum critical theory is generically described by a 
black hole spacetime. Within the framework of linear response, 
one analyses the linear perturbations about the black hole 
that are induced by applied electric fields
and thermal gradients. By solving the linear equations for the perturbations
in the bulk spacetime and then examining their behaviour
at the boundary one can extract the electric
and heat currents that are produced by the applied sources and hence obtain the matrix of thermoelectric conductivities.

It has recently been shown, quite generally, that the calculation of the DC conductivity actually simplifies dramatically and
boils down to solving a generalised system of Stokes equations on the black hole event horizon 
\cite{Donos:2015gia,Banks:2015wha}.
For some simple classes of black hole solutions the Stokes equations
can be solved explicitly in terms of the behaviour of the black hole solution at the horizon and one
recovers the results obtained in 
\cite{Iqbal:2008by,Donos:2014uba,Donos:2014cya,Donos:2014yya}, thus placing them in a more general context. 
The black holes considered in \cite{Donos:2015gia,Banks:2015wha}
were electrically charged and static solutions of Einstein-Maxwell-scalar
theory in $D\ge 4$ spacetime dimensions and allowed for general spatial inhomogeneities at the AdS boundary.
Such inhomogeneities, which break translation invariance explicitly, 
provide a mechanism for momentum to dissipate \cite{Horowitz:2012ky} 
and are, generically, essential
in order to obtain finite DC responses. 
A particularly interesting aspect of the inclusion of bulk scalar fields was the 
appearance of a novel viscous term in the Stokes equations on the black hole horizon.

The Stokes equations can be used to obtain some important insights about transport in the holographic setting.
For example, some interesting results can be obtained for the 
class of black hole solutions which have horizons that arise as perturbative expansions about flat horizons. 
Indeed explicit expressions for the leading order behaviour of the conductivity were obtained in
\cite{Donos:2015gia,Banks:2015wha} by solving the Stokes equations perturbatively. 
In particular, while the Wiedemann-Franz law is
violated, it was shown
that to leading order $\bar\kappa(\sigma T)^{-1}=s^2/\rho^2$, where $s$ and $\rho$ are the entropy density and the charge density, respectively. One interesting application is in
the context of ``coherent" or ``good" metallic ground states whose low temperature behaviour is governed by a translationally invariant ground state with momentum dissipation
being associated with RG irrelevant operators, with the latter governing the perturbative expansion. The results
in \cite{Donos:2015gia,Banks:2015wha} complement those obtained using the memory matrix formalism
\cite{Hartnoll:2012rj,Mahajan:2013cja}. 
Another application is in the context of perturbative lattices i.e. where the UV deformation
is a small perturbation about a translationally invariant solution. Related results were also
obtained from a hydrodynamic point of view in \cite{Lucas:2015sya}. Other work on DC conductivity in a 
hydrodynamic limit includes \cite{Hartnoll:2007ih,Davison:2014lua,Davison:2015bea,Blake:2015epa,Lucas:2015lna,Blake:2015hxa}.
In a slightly different direction, the Stokes equations
have been used to show that Einstein-Maxwell theory in $D=4$, with some additional assumptions, 
cannot exhibit metal-insulator transitions \cite{Grozdanov:2015qia}. This result is consistent with the fact that such 
transitions have been observed in systems that have extra matter fields or higher dimensions \cite{Donos:2012js,Donos:2014uba,Gouteraux:2014hca}.

The principal purpose of this paper is to extend the results for the DC conductivity in 
\cite{Donos:2015gia,Banks:2015wha}
from static black holes to stationary black holes that can also include spatially varying 
magnetic fields\footnote{Conductivities for black holes with constant magnetic fields were
studied without lattices in e.g. \cite{Hartnoll:2007ai,Hartnoll:2007ip,Lindgren:2015lia} and for Q-lattices in 
\cite{Blake:2014yla,Blake:2015ina}. Results in the context of massive gravity were reported on 
in \cite{Amoretti:2015gna}.}.
In particular, we will 
consider black hole solutions in which the asymptotic behaviour of the spatial components of the bulk gauge-field, $A_i$,
at the holographic boundary are associated with magnetic field sources which include
constant as well as spatially varying pieces. The latter provide
sources for the electric currents in the dual field theory and while they
have not been considered much in the holographic literature (an example is
\cite{Donos:2012wi}), we
advocate here that they are an interesting framework for considering holographic duals of magnetic impurities.

In the dual field theory these deformations break time reversal invariance and so the background 
black holes can have non-vanishing local electric current densities, $J^{(B)i}_\infty$,  
and heat current densities, $Q^{(B)i}_\infty$,  where $i$ is a spatial index and
the superscript $(B)$ refers to the background black hole solution.
We will assume that the black holes have Killing horizons which corresponds to the dual
field theory being in thermal equilibrium. In general, thermal equilibrium and the Ward identities
imply\footnote{The result for $J$ immediately follows from the fact that time derivatives vanish
in thermal equilibrium. The result for $Q$ follows after contracting the Ward identity for diffeomorphisms with
the vector $k=\partial_t$.} that the current densities are conserved: $\partial_iJ^{(B)i}_\infty=\partial_iQ^{(B)i}_\infty=0$.
Furthermore, since momentum relaxes in the backgrounds that we consider
there will be no net current fluxes (defined in section \ref{usersguide})
and we deduce that the dual field theories can have currents which are
magnetisation currents of the form $J^{(B)i}_\infty=\partial_j M^{(B)ij}$,
$Q^{(B)i}_\infty=\partial_j M^{(B)ij}_T$ with $M^{(B)ij}=-M^{(B)ji}$ and  $M^{(B)ij}_T=-M^{(B)ji}_T$.
Explicit expressions for the local magnetisation density, $M^{(B)ij}$, and the
local thermal magnetisation density, $M^{(B)ij}_T$, in terms of the bulk fields of the background solution will be given in section \ref{currentsec} 
(and also section \ref{currentsmarktwo}).
It is also worth mentioning that black holes with such magnetisation currents also arise
at finite charge density in the context of phases in which translations and time reversal invariance
are broken spontaneously e.g. \cite{Nakamura:2009tf,Donos:2011bh,Donos:2011qt,Ammon:2011je,Donos:2012wi,Donos:2013wia,
Withers:2013loa,Rozali:2013ama,Withers:2014sja}.

In addition to the asymptotic behaviour of the gauge-field $A_i$ described above, we will
also include analogous sources for the off-diagonal  
components of the metic $g_{ti}$, where $t$ is the time coordinate. The analogue of a
constant magnetic field is now a kind of NUT charge. In appendix \ref{nutchge} we discuss a simple
analytic black hole solution with such a charge, but we note that it has closed time like curves.
While the physical role of the NUT charges, in general, is a little unclear, our analysis will nevertheless include them.
On the other hand, again analogous to the gauge-field, the asymptotic behaviour of $g_{ti}$ will also allow for local
sources for the heat currents, which appear to be interesting deformations worthy of further exploration. 
All of these deformations break time-reversal invariance and will, in general, also give rise to magnetisation electric and heat currents.

We will show that the DC conductivity for this very general class of black holes can be obtained from a generalised Stokes
flow at the black hole horizon. The Stokes equations contain some extra terms that are analogous to those appearing in magneto hydrodynamics\footnote{In a different setting non-linear magneto hydrodynamics was related to solutions of
Einstein equations in \cite{Lysov:2013jsa}, extending \cite{Bredberg:2011jq}.} .
 There is also an additional viscous term arising, roughly speaking, from the behaviour
of $g_{ti}$ at the horizon. These new generalised Stokes equations retain many of the
properties proved in \cite{Donos:2015gia,Banks:2015wha} for the static, electrically charged black holes. One difference, however, is that we have not 
managed to derive them from a variational principle.

In generalising the results on DC conductivity to this more general class of black holes there is an
important subtlety. In the presence of magnetic fields and magnetisation currents one needs to carefully
identify the correct transport currents. A general discussion of this issue was first made in
\cite{PhysRevB.55.2344}. In the holographic context, in the special case of constant magnetic fields 
and for special classes of black holes, the precise transport currents were identified in 
\cite{Blake:2015ina,Hartnoll:2007ih}. Here we will make a more general identification.

In section \ref{sec2} of this paper we introduce the general class of black hole solutions that we shall
be considering. Section \ref{usersguide} is a simple ``users guide" that summarises the computational 
procedure for obtaining the DC conductivity of the dual field theory by solving a precise set of 
generalised Stokes equations on the black hole horizon.
This section also includes a brief discussion of some of the properties of the Stokes equations.
In section \ref{sec4} and \ref{der2} we provide two derivations of the results explained in section \ref{usersguide}.

In section \ref{sec4} we carry out the analysis for black hole solutions for which there is a single horizon and
in addition it is possible to choose coordinates $(t,r,x^i)$ globally outside the horizon. Furthermore,
we often consider the case when all fields depend periodically on the spatial coordinates $x^i$, in which case
they effectively parametrise a torus.
One reason for considering this class
is that some readers might find the analysis simpler to follow, especially with a globally defined radial coordinate. 
The second
reason is that this is the class of black hole solutions that are
most amenable to being constructed by numerical techniques. 
In section \ref{der2} we use the language of differential forms to derive our main results for general black hole
solutions, including the possibility that there is more than one event horizon.
For readers familiar with the language of forms
the derivation of section \ref{der2} is in many ways simpler than the derivation in section \ref{sec4}.

In section \ref{exsect} we analyse the Stokes equations of section \ref{usersguide} for special classes of black hole solutions
where we can solve the equations explicitly in terms of the behaviour of the solution
at the black hole horizon. We consider general Q-lattice black holes \cite{Donos:2013eha} with constant 
magnetic fields. As a special sub-case we recover the result of \cite{Blake:2015ina}
as well as the older result for the electric DC conductivity for the dyonic AdS-RN black brane
in $D=4$ \cite{Hartnoll:2007ai}. We also show that in the case of black hole solutions
which only depend on one of the spatial directions, the Stokes equations can be explicitly solved in closed form.

We briefly conclude in section \ref{sec7}. The paper also contains various material in six appendices. In particular,
appendix \ref{sduality} explores the interesting feature,
for the special case of holographic models in $D=4$, that the generalised Stokes equations have
an S-duality invariance, independently of whether the bulk theory itself exhibits S-duality.

\section{The background black holes}\label{sec2}

We will consider theories in $D$ spacetime dimensions which couple the metric, $g$,
to a gauge-field, $A$, and a single scalar field, $\phi$. The extension 
to include additional scalar fields is straightforward as we discuss later. 
We focus on $D\ge 4$ and
the action is given by
\begin{align}\label{eq:bulk_action}
S=\int d^D x \sqrt{-g}\,\left(R-V(\phi)-\frac{Z(\phi)}{4}\,F^{2}-\frac{1}{2}\left(\partial\phi \right)^2\right)\,.
\end{align}
We will assume that $V(0)=-(D-1)(D-2)$ and $V'(0)=0$. 
This ensures that a unit radius $AdS_{D}$ solves
the equations of motion with $\phi=0$ and this is dual to a CFT with a stress tensor,
dual to the metric, a global $U(1)$ current, dual to $A$, and an additional operator ${\cal O}$ dual to $\phi$.
Note that we have set $16\pi G=1$, as well as setting the AdS radius to unity, for convenience. 
 
Our analysis will cover a very general class of stationary black hole solutions with Killing vector $k$. The solutions asymptotically approach $AdS_D$ and allow for very general spatially dependent (lattice) deformations as well as non-vanishing magnetic fields. We will assume that the Lie derivatives of $A$ and $\phi$ with respect to the Killing vector all vanish: $L_kA=L_k\phi=0$. 
Introducing local coordinates $x^\mu=(t,r,x^i)$
at the asymptotic boundary, with $k=\partial_t$, we assume that as $r\to\infty$ we have
\begin{align}\label{asmet}
ds^2&\to r^{-2}dr^2+r^2\left[g^{(\infty)}_{tt}dt^2+g^{(\infty)}_{ij}dx^idx^j+2g_{ti}^{(\infty)}dtdx^i\right]\,,\nn
A&\to A^{(\infty)}_t dt+A^{(\infty)}_i dx^i\,, \nn
\phi&\to r^{\Delta-D+1}\phi^{(\infty)}\,,
\end{align}
where $g^{(\infty)}_{tt}$ etc. are functions of the spatial coordinates, $x^i$, only. The $x^i$ parametrise a $d$-dimensional manifold, $\Sigma_d$, with $d=D-2$, that can have arbitrary topology.
A subclass of special interest is when all deformations are periodic in the coordinates $x^i$ and, effectively, we can then consider the $x^i$ to parametrise a torus, $\Sigma_d=T^d$. Notice that \eqref{asmet} allows for a general lattice deformation of the CFT.  In particular, we have allowed for a non-trivial spatially modulated boundary metric parametrised by
$g^{(\infty)}_{tt}$, $g^{(\infty)}_{ij}$, $g_{ti}^{(\infty)}$. A non-trivial $\phi^{(\infty)}$ is associated with a spatially modulated 
deformation of the CFT by the operator ${\cal O}$, which we have assumed has scaling dimension $\Delta\le D-1$, corresponding to a relevant or marginal operator. 
Next, $A^{(\infty)}_t$ parameterises a spatially modulated chemical
potential which can, of course, include a constant piece as well as pieces that depend on the spatial coordinates. 
Finally, $A^{(\infty)}_i$ allows for constant and spatially modulated magnetic fields.
For example, in the case that the coordinates $x^i$ parametrise a torus we can
consider $A^{(\infty)}_idx^i = a_i dx^i + f_{ij}x^idx^j$ where the components $a_i$ are periodic in the $x^i$ and parametrise sources for the currents, or equivalently spatially dependent local magnetic fields, while
the $f_{ij}$ are constants and parametrise the constant magnetic fields.
Note that there is an analogous situation with $g_{ti}^{(\infty)}$: the periodic
components are associated with sources for the spatial momentum while the non periodic constant 
components give what one might call ``NUT charges". As somewhat of an aside we present an analytic solution with constant NUT charge
in appendix \ref{nutchge}, but we note that it has closed time-like curves.

We now discuss the behaviour at the black hole horizons, 
allowing for the possibility that there is more
than one\footnote{Examples of such solutions
have been discussed in \cite{Horowitz:2014gva}.}. We will assume that each horizon is a Killing horizon with respect to $k$. In particular,
the Killing vector $k$ is time-like outside all of the horizons and becomes null at each horizon and furthermore, all of the black holes will have the same temperature $T$. We also assume that it is possible to introduce a 
a time coordinate, $t$, that is globally defined outside the event horizons, with $k=\partial_t$.
In order to describe the solution at one of the horizons we introduce local coordinates $x^\mu=(t,r,x^i)$ and consider the general expressions
\begin{align}\label{met}
ds^2&=g_{tt}dt^2+g_{rr}dr^2+g_{ij}dx^idx^j+2g_{tr}dt dr+2g_{ir}dx^idr+2g_{ti}dx^idt\,,\nn
A&=A_tdt+A_r dr+A_i dx^i\, .
\end{align}
We assume that the black hole horizon is located at $r=0$ and introduce 
a function $U(r)$ that is analytic at $r=0$, $U\left(r\right)=r\left(4\pi\,T+\dots\right)$. We demand that the black hole horizon is regular after Wick rotating $t=-i\tau$ and then employing
the cartesian coordinates $X=\sqrt{r}\cos(2\pi T\tau)$, $Y=\sqrt{r}\sin(2\pi T\tau)$. 
We conclude that as $r\to 0$ we have:
\begin{align}\label{defhqs}
g_{tt}(r,x)&=-U(G^{(0)}(x)+...)\,,\qquad
g_{rr}(r,x)=U^{-1}(G^{(0)}(x)+...)\,,\nn
g_{tr}(r,x)&=U({g_{tr}}^{(0)}(x)+...)\,,\qquad
g_{ti}(r,x)=U({G^{(0)}(x)}{\chi_i}^{(0)}(x)+...)\,,\nn
A_t(r,x)&=U(\frac{G^{(0)}(x)}{4\pi T}A_t^{(0)}(x)+...)\,,
\end{align}
where the dots refer to higher powers in $r$ and all other quantities are, in general, non-vanishing at the horizon:
\begin{align}\label{defhqs2}
g_{ij}(r,x)&=h_{ij}^{(0)}(x)+ ...\,,\quad
g_{ir}(r,x)=g_{ir}^{(0)}(x)+...\,,\cr
A_i(r,x)&=A_i^{(0)}(x)+...\,,\,\,
A_r(r,x)=A_r^{(0)}(x)+...\,,\,\,
\phi(r,x)=\phi^{(0)}(x)+...\,.
\end{align}
Note that the extra factors of ${G^{(0)}(x)}$ in the last two equations in \eqref{defhqs} have been added
for later convenience.

A class of solutions of particular interest is when there is a single black hole horizon and, in addition, the coordinates 
$(t,r,x^i)$ are valid all the way from just outside the horizon to the $AdS$ boundary (i.e. the coordinates in
\eqref{asmet} and \eqref{met} are the same). In particular, the horizon manifold will be the
same as the spatial sections at the asymptotic boundary, i.e. $\Sigma_d$, but in general, $h_{ij}^{(0)}\ne g_{ij}^{(\infty)}$.
For ease of presentation we will sometimes use this class
of solutions to explain some results, but we emphasise that this is just for convenience.

\section{A users guide to DC conductivity}\label{usersguide}
In this section we will summarise the practical procedure for calculating
the thermoelectric DC conductivity in terms of the behaviour of the black hole solution at
the horizon. Specifically, it is necessary to solve a system of ``generalised Stokes equations"
on the black hole horizon.
For simplicity, here we will discuss the class of black holes that we mentioned at the end
of the last section in which there is a single black hole horizon and the coordinates \eqref{met} are valid
everywhere outside the horizon. Furthermore, we will focus, at some point below, on the case of periodic lattices for which
the spatial manifold $\Sigma_d$ can be assumed to be a $d$-dimensional torus. In subsequent sections we will derive these results, explaining the origin of the fluid equations on the horizon, as well
as generalising away from this class. In section \ref{exsect} we analyse some concrete examples. At the end of this section we will describe some general properties of the
generalised Stokes equations.

We introduce one-forms $E=E_i(x)dx^i$ and $\zeta=\zeta_i(x)dx^i$, defined on a general $\Sigma_d$,  
which parametrise the applied external electric field and thermal gradients, and we demand that they are closed: 
$dE=d\zeta=0$. 
One then must solve the
following system of generalised Stokes equations for the variables $(v^i,w,p)$ on the black hole horizon $\Sigma_d$:
\begin{align}
-2{\nabla}^j{\nabla}_{(i}v_{j)}+v^j[{\nabla}_j\phi^{(0)}{\nabla}_i\phi^{(0)}&+4\pi Td{\chi}^{(0)}_{ji}]
-{F}_{ij}^{(0)}\frac{{J^j_{(0)}} }{\sqrt{h^{(0)}}}
\nn&=Z^{(0)}A_t^{(0)}(E_i+{\nabla}_iw)+4\pi T\zeta_i-{\nabla}_ip\,,\nn
{\nabla}_iv^i&=0\,,\qquad\qquad
\partial_iJ^i_{(0)} =0\,,
\label{constrainteqns}
\end{align}
where 
\begin{align}\label{expforJ}
J^i_{(0)} \equiv \sqrt{h^{(0)}}Z^{(0)}(E^i+{\nabla}^iw+A_t^{(0)}v^i+{F}^{i(0)}_{\,\,\,\,k}v^k)\,.
\end{align}
Here, the covariant derivative, $\nabla$, is defined with respect to the horizon metric $h_{ij}^{(0)}$ and all indices are raised and lowered
with respect to this metric. The horizon quantities for the background black holes, $\phi^{(0)}, {\chi}^{(0)}, A_t^{(0)}$, are defined in \eqref{defhqs},\eqref{defhqs2} and ${F}_{ij}^{(0)}\equiv \partial_i{A_j}^{(0)}-\partial_j{A_i}^{(0)}$. We have also defined
$Z^{(0)}\equiv Z(\phi^{(0)})$. Compared to the results obtained for the purely electrically charged black holes
in \cite{Donos:2015gia,Banks:2015wha} we highlight
 the term ${F}_{ij}^{(0)}{{J^j_{(0)}}}$, which perhaps might have been expected by analogy with magneto hydrodynamics, 
as well as the novel viscous term\footnote{It would be interesting to see if similar terms can also appear in the non-linear
Navier-Stokes equations that arise in different gravitational calculations, such as 
\cite{Bhattacharyya:2008kq,Fouxon:2008tb,Eling:2009pb,Bredberg:2011jq}.}
involving $d{\chi}^{(0)}_{ij}\equiv \partial_i{\chi}^{(0)}_j-\partial_j{\chi}^{(0)}_i$. It is worth emphasising that 
$d{\chi}^{(0)}$ can, in principle, be non-zero even if $g_{ti}^{(\infty)}=0$ in \eqref{asmet}. An illuminating equivalent
presentation of these equations is given in \eqref{auxstokes}, below.

Having solved these linear partial differential equations for $(v^i,w,p)$, we can obtain expressions for $J^i_{(0)} $ and $Q^i_{(0)}$, the electric and heat current densities at the horizon, respectively,
in terms of the sources $E_i,\zeta_i$, where
\begin{align}
Q^i_{(0)} \equiv 4\pi T\sqrt{h^{(0)}}v^i\,.
\end{align}
 To obtain the DC conductivity we need to obtain  suitable current flux densities at the asymptotic boundary from this horizon data.

To do this we now focus on solutions which are periodic in the $x^i$ coordinates with period $L_i$: $x^i\sim x^i+L_i$ and we illustrate for the cases of $d=2,3$. 
For $d=2$ (i.e. $D=4$) we define the following current flux densities at the horizon
\begin{align}\label{avecur}
\bar J^1_{(0)}\equiv \frac{1}{ L_2}\int J^1_{(0)}  dx^2\,,\qquad
\bar J^2_{(0)}\equiv \frac{1}{ L_1}\int J^2_{(0)}  dx^1\,,
\end{align}
where $\bar J^1$ and $\bar J^2$ is the current flux density through the $x^2$ and $x^1$ planes, respectively, and
we define $\bar Q^i$ in a similar way. 
Similarly, for $d=3$ (i.e. $D=5$)
\begin{align}\label{avecur2}
\bar J^1_{(0)}\equiv \frac{1}{ L_2L_3}\int  dx^2dx^3 J^1_{(0)} \,,\quad
\bar J^2_{(0)}\equiv -\frac{1}{ L_3L_1}\int  dx^1dx^3 J^2_{(0)} \,,\quad
\bar J^3_{(0)}\equiv \frac{1}{ L_1L_2}\int  dx^1dx^2 J^3_{(0)} \,.
\end{align}
It is an important fact that these current fluxes $\bar J^i_{(0)}$ are constants. For example, $\bar J^1_{(0)}$ in \eqref{avecur}
is clearly independent of $x^2$. In addition, the derivative with respect to $x^1$ can be seen to vanish after
using $\partial_iJ^i_{(0)}=0$ in the integrand. These fluxes could thus also be defined by doing an averaged integral over all spatial directions.

For the sources we write 
\begin{align}\label{ees}
E=\bar E_i dx^i+de\,,\qquad \zeta=\bar\zeta_i dx^i+ dz\,,
\end{align} with constant $\bar E_i,\bar \zeta_i$ and
$e,z$ arbitrary periodic functions. 
The thermoelectric DC conductivity  matrix is then obtained via
\begin{align}\label{bigform2}
\left(
\begin{array}{c}
\bar J^i_{(0)}\\\bar Q^i_{(0)}
\end{array}
\right)=
\left(\begin{array}{cc}
\sigma^{ij} & T\alpha^{ij} \\
T\bar\alpha^{ij} &T\bar\kappa^{ij}   \\
\end{array}\right)
\left(
\begin{array}{c}
\bar E_j\\ \bar\zeta_j
\end{array}
\right)\,.
\end{align}

If this is in fact the thermoelectric DC conductivity, as claimed, we must be able to identify the sources and
the current flux densities at the horizon with those at the asymptotic boundary. For the sources (in the present context) this is trivial because $\bar E_i,\bar \zeta_i$ are constant. For the currents, however, there is an important subtlety in the presence of magnetic fields \cite{PhysRevB.55.2344,Hartnoll:2007ih,Blake:2015ina}. In order to extract the physical transport currents that couple to the external sources, we must subtract out the magnetisation currents. 
As mentioned earlier, the holographic current densities of the background
black holes, $J^{(B)i}_\infty$, $Q^{(B)i}_\infty$,
are magnetisation currents of the form
\begin{align}\label{jayback0}
J^{(B)i}_\infty=\partial_j M^{(B)ij}\,,\qquad
Q^{(B)i}_\infty=\partial_j M^{(B)ij}_T\,,
\end{align}
where explicit expressions for the local magnetisation density, $M^{(B)ij}$, and the
local thermal magnetisation density, $M^{(B)ij}_T$, in terms of the bulk fields of the background solution will be given in section \ref{currentsec}.
In particular the background black holes have
vanishing current flux densities: $\bar J^{(B)i}_\infty=\bar Q^{(B)i}_\infty=0$. 
The constant ``transport current flux densities" of the dual field theory, 
$\bar{\mathcal{J}}^i$, $\bar{\mathcal{Q}}^i$,
are then defined via 
\begin{align}\label{initialbarj}
\bar{\mathcal{J}}^i&=\bar J^i_{\infty}+\overline{M^{{(B)}ij}\zeta_j},\cr
\bar{\mathcal{Q}}^i&=\bar Q^i_{\infty}+\overline{M^{(B)ij} E_j}+2\overline{M^{(B)ij}_{T}\zeta_j}\,,
\end{align}
where $J^i_{\infty}, Q^i_{\infty}$ are the total holographic currents of the dual field theory\footnote{There is a small subtlety
concerning counterterms which we will come back to later. We emphasise, though, that it does not affect 
$\bar{\mathcal{J}}^i$ and $\bar{\mathcal{Q}}^i$.}.
A key fact
is that the transport current flux densities have exactly the same value as the current flux densities at the horizon:
\begin{align}
\bar{\mathcal{J}}^i_\infty=\bar J^i_{(0)} \,,\qquad
\bar{\mathcal{Q}}^i_\infty =\bar Q^i_{(0)}\,,
\end{align}
and hence \eqref{bigform2} is relating the transport current flux densities of the dual field theory
to the sources which defines
the DC conductivity matrix.

\subsection{Some properties of the Stokes equations}\label{genpropsstokes}
The generalised Stokes equations \eqref{constrainteqns} satisfy various properties, for general $\Sigma_d$, 
that were shown to hold in the case of
purely electrically charged, static black holes in \cite{Donos:2015gia,Banks:2015wha}. For example, if we multiply the first of the 
Stokes equations \eqref{constrainteqns} by $v^j$ and then integrate over the horizon
we obtain
\begin{align}\label{power}
\int d^dx\sqrt{h^{(0)}}\Big(2{\nabla}^{(i}v^{j)}{\nabla}_{(i}v_{j)}+[v^j{\nabla}_j\phi^{(0)}]^2+Z^{(0)}|&E_i+{\nabla}_iw+{F}^{(0)}_{ij}v^j|^2\Big)\nn
&=\int d^dx ({J^i_{(0)}}E_i+{Q^i_{(0)}}\zeta_i)\,.
\end{align}
The left hand side is a positive quantity which implies the positivity of the thermoelectric conductivity matrix. 

We next consider the issue of uniqueness of the solutions to the Stokes equations with the same $E,\zeta$. 
Taking the difference of two such solutions we obtain a solution to the equations with $E=\zeta=0$. Denoting such
a solution again by $v,w,p$, we easily deduce from \eqref{power} that we must have
${\nabla}^{(i}v^{j)}=0$, $v^i{\nabla}_i\phi^{(0)}=0$, ${\nabla}_iw+v^j{F}_{ij}^{(0)}=0$. From
\eqref{constrainteqns} we then deduce that ${\nabla}_ip+(4\pi T)v^jd{\chi}_{ji}^{(0)}=0$, 
and $v^i\nabla_i A^{(0)}_t$=0. Thus, we must have, in particular, that
$v^i$ is a Killing vector which preserves $w,p$ and 
$A^{(0)}_t$, but we note that in general this smaller set of conditions are not sufficient for finding a solution to the source free equations\footnote{For example, consider the standard $D=4$ dyonic black hole given in \eqref{adsexact}. The vector $v^i=(1,0)$ is Killing, but we cannot just demand that $w$ is independent of $x$ we also need $\partial_y w=B$.}.
A solution to the source free equations on the horizon, supplemented by the rest of the perturbation in the bulk,
is in fact a quasi-normal zero mode of the background black hole.
If the black hole admits such a zero mode then the DC conductivity is not defined\footnote{We expect that
there are no solutions to the sourced Stokes equations if and only if a zero mode exists.}.

As in \cite{Banks:2015wha} we can eliminate the sources locally from the equations but not globally. Indeed, locally, we can write
$E=d\tilde e$ and $\zeta= d\tilde z$, for locally defined functions $\tilde e,\tilde z$, which we can then eliminate by redefining $w,p$, respectively. This also shows that the exact part of the forms $E,\zeta$ does not contribute to the DC conductivity, and this justifies why we ignored the exact terms in \eqref{ees} in obtaining the DC conductivity above.

Unlike the case of static black holes with purely electric fields of \cite{Banks:2015wha}, when $d\chi^{(0)}\ne0$, ${F}^{(0)}_{ij}\ne 0$  we have not found a way to obtain the generalised Stokes equations \eqref{constrainteqns} from a variational principle. 
For the static black holes the variational principle was used to argue that Onsager relations were always valid. In general,
the Onsager relations imply that the transpose of the full matrix appearing in \eqref{bigform2} is symmetric provided
that one replaces the source terms with their time reversed quantities. For the black holes of this paper we need to reverse the sign of
$\chi_{i}^{(\infty)}$ and $A_i^{(\infty)}$. Since we have assumed that the time coordinate $t$ is globally defined (outside the event horizons), this corresponds
to reversing the sign of $\chi^{(0)}$ and $A_i^{(0)}$ at the horizon. For particular classes of black holes, discussed in section \ref{exsect}, we will show
that the Onsager relations are indeed satisfied. We leave a general proof to future work.

It is also interesting to write the Stokes equations at the horizon in a way that further highlights
their structure as equations for an auxiliary fluid at the horizon. The current densities at the horizon can be written
\begin{align}\label{eq:JQ_hor2}
J^i_{(0)} &=\rho_H v^i+\Sigma_H^{ij}\left(\partial_j w+E_j+{F}^{(0)}_{jk}v^k\right)\,,\nn
Q^i_{(0)} &=Ts_Hv^i\,,
\end{align}
where we define the horizon quantities
\begin{align}\label{Hquants}
\rho_H&=\sqrt{h^{(0)}}Z^{(0)}A_t^{(0)},\qquad s_H=4\pi \sqrt{h^{(0)}}\,,\nn
\Sigma^{ij}_H&=\sqrt{h^{(0)}}Z^{(0)}h^{ij}_{(0)},\qquad \eta_H=\frac{s_H}{4\pi}\,.
\end{align}
The generalised Stokes equations are then
\begin{align}\label{auxstokes}
\eta_H\Big(-2{\nabla}^j{\nabla}_{(i}v_{j)}+v^j&{\nabla}_j\phi^{(0)}{\nabla}_i\phi^{(0)}\Big)
-d{\chi}^{(0)}_{ij}{Q^j_{(0)}}-{F}_{ij}^{(0)}{J^j_{(0)}}
\nn&=\rho_H(E_i+{\nabla}_iw)+Ts_H\left(\zeta_i-{\nabla}_i\frac{p}{4\pi T}\right)
 \nn
\partial_i Q^i_{(0)}&=0\,,\qquad\qquad
\partial_iJ^i_{(0)} =0\,,
\end{align}
In \eqref{eq:JQ_hor2} we see that $\rho_H$, $s_H$ and $\Sigma_H^{ij}$ are local coefficients
in constitutive relations for the auxiliary fluid. We emphasise that, in general, $\Sigma_H^{ij}$
is not the same as the physical electrical conductivity $\sigma^{ij}$ of the dual field theory obtained
via the horizon calculation \eqref{bigform2}. However, for some special sub-classes of black holes
they happen happen to be the 
same\footnote{Furthermore $\Sigma_H$ is sometimes the same
as the electrical conductivity
at zero heat flow, $\sigma_{Q=0}\equiv  \sigma-T\alpha\bar\kappa^{-1}\bar\alpha$, but again they
differ in general.}.

It is also worth noting that similar comments apply to $\eta_H$. In fact, since the background black holes
we are considering can be anisotropic, the holographic shear viscosity is actually a tensor and clearly is not directly related
to $\eta_H$. On the other hand, the total charge density, $\rho$, and the entropy density, $s$, of the dual CFT 
can be obtained by averaging $\rho_H$ and $s_H$ over the torus, respectively.

As noted earlier, we have set $16\pi G=1$. Reinstating factors of $G$ we would find
$\eta_H,\rho_H,s_H,J^i_{(0)}$ are all rescaled by $1/(16\pi G)$, consistent with \eqref{auxstokes}. This also
leads to an overall factor of $1/(16\pi G)$ appearing in front of the conductivity matrix in \eqref{bigform2},
from which one can deduce the overall dependence on $N$, the number of colours of the underlying dual CFT.

For the special case of $D=4$ spacetime dimensions the Stokes equations have an S-duality invariance, independent
of whether or not the bulk equations of motion are S-duality invariant. The implications for the DC conductivity are explored in
appendix \ref{sduality}.

\section{Derivation for a subclass of black hole solutions}\label{sec4}
In this section we will explain how the results for calculating the DC conductivity summarised in
the previous section can be derived. As in that section we will assume that there is
a single black hole and the coordinates $(t,r,x^i)$ are valid everywhere outside the horizon. Similarly, for the most part we will allow for arbitrary $\Sigma_d$. 
We will generalise to other cases in section \ref{der2} using a more general formalism.

\subsection{Perturbing the black holes}
In order to introduce suitable sources for the electric and heat currents we consider the following linear perturbation
of the black hole solution \eqref{met}:
\begin{align}
\delta(ds^2)&=\delta g_{\mu\nu}dx^\mu dx^\nu-2tg_{tt}\zeta_i dt dx^i+t(g_{ti}\zeta_j+g_{tj}\zeta_i)dx^i dx^j+2tg_{tr}\zeta_idr dx^i\,,\nn
\delta A&=\delta a_\mu dx^\mu-t E_i dx^i+tA_t\zeta_i dx^i\,,\label{heatpertansatz}
\end{align}
as well as $\delta\phi$. The source terms $E=E_i(x)dx^i$ and $\zeta=\zeta_i(x)dx^i$ are one-forms on $\Sigma_d$, but the other functions in the perturbation depend on both $r$ and $x^i$. We impose that $E$ and $\zeta$ are closed forms, $dE=d\zeta=0$. This generalises what was studied in \cite{Donos:2015gia,Banks:2015wha} and in particular all
time dependence of the equations of motion is satisfied, to linear order in the perturbation\footnote{To see this, note that since $E$ and $\zeta$ are closed, locally we can write $E=de$ and $\zeta=dz$. First make the gauge transformation $\delta A=d(te)=edt+t E_idx^i$. This removes the time-dependent $t E_i dx^i$ term from (\ref{heatpertansatz}). Then make the coordinate transformation $t\rightarrow t(1-z)$ so that $dt \rightarrow dt (1-z)- t\zeta_i dx^i$. This removes the remaining time dependence present in (\ref{heatpertansatz}) (to linear order in the perturbation).}.

At the AdS boundary we assume that the fall-off of the perturbation as $r\to\infty$ is such that
the only sources are given by $E$ and $\zeta$. We do not need to make these conditions more explicit to obtain our results.

It is important that the perturbation is regular at the black hole horizon. Switching from the time coordinate $t$ to the
Kruskal coordinate $v=t+\frac{\ln r}{4\pi T}+\dots$, near $r=0$
we demand that we have:
\begin{align}
\delta g_{tt}&=U(\delta g_{tt}^{(0)}+...)\,,\qquad
\delta g_{rr}={U}^{-1}(\delta g_{rr}^{(0)}+...)\,,\cr
\delta g_{rt}&=\delta g_{rt}^{(0)}+...\,,\qquad
\quad\delta g_{ti}=\delta g_{ti}^{(0)}+...\,,\cr
\delta g_{ij}&=\delta g_{ij}^{(0)}+...\,,\qquad
\quad\delta g_{ri}={U}^{-1}(\delta g_{ti}^{(0)}+...)\,,
\end{align}
where $\delta g_{tt}^{(0)}+\delta g_{rr}^{(0)}-2\delta g_{rt}^{(0)}=0$. Also:
\begin{align}
\delta a_t&=\delta a_t^{(0)}(x)+...\,,\qquad\qquad\qquad
\delta a_r={U}^{-1}(\delta a_t^{(0)}(x)+...)\,,\cr
\delta a_j&=\frac{\ln r}{4\pi T}(-E_j+A_t\zeta_j)+...\,,\qquad\delta\phi=\delta\phi^{(0)}+...\,.
\end{align}

\subsection{Constraints at the horizon}\label{conhor}
We have assumed (in this section) that the black hole solution admits a global foliation
by surfaces with constant $r$. Thus we can rewrite the equations of motion using a radial Hamiltonian decomposition. To do this we introduce the normal vector $n^\mu$, satisfying $n^\mu n_\mu=1$.
The $D$-dimensional metric $g_{\mu\nu}$ induces a $(D-1)$-dimensional Lorentzian 
metric on the slices of constant $r$ via $h_{\mu\nu}=g_{\mu\nu}-n_\mu n_\nu$.
The lapse and shift vectors are given by $n_\mu=N(dr)_\mu$ and $N^\mu=h^\mu{}_\nu r^\nu$, respectively, where
$r^\mu=(\partial_r)^\mu$. The gauge-field components are decomposed via 
\begin{align}
b_\mu=h_\mu{}^\nu A_\nu,\qquad
\Phi=-Nn^\mu A_\mu\,,
\end{align}
and we define $f_{\mu\nu}=\partial_\mu b_\nu-\partial_\nu b_\mu$. 
The momenta conjugate to $h_{\mu\nu}$, ${b}_{\mu}$ and $\phi$ are given by
\begin{align}\label{momdefns}
\pi^{\mu\nu}&=-\sqrt{-h}\,\left( K^{\mu\nu}-K\,h^{\mu\nu}\right)\,,\notag\\
\pi^\mu&=\sqrt{-h} Z F^{\mu\rho}n_\rho\,,\nn
\pi_\phi&
=-\sqrt{-h}n^\nu\partial_\nu\phi\,,
\end{align}
respectively, where $K_{\mu\nu}=\frac{1}{2}{\mathcal L}_n h_{\mu\nu}$. As usual the Hamiltonian density can be written as a sum of constraints
\begin{align}
\mathcal{H}&=
N\,H+N_{\mu}\,H^{\mu}+\Phi\,C \,.
\end{align}
Explicit expressions for $H$, $H_\mu$ and $C$
in terms of canonical variables were given in equations (A.9)-(A.11) of \cite{Banks:2015wha}.

As in \cite{Donos:2015gia,Banks:2015wha}, for the perturbed metric we want to evaluate the constraints on a surface of constant $r$ close to the horizon and then take the limit $r\to 0$. The calculations, which are rather lengthy, closely follow those in \cite{Banks:2015wha}. For the
background black hole solution we find that the constraints are all satisfied on the horizon. At linear order in the
perturbation we find that the Gauss Law constraint $C=0$ implies $\partial_i\pi^i=0$
on the horizon, which is the equation $\partial_iJ^i_{(0)}=0$ given in the generalised Stokes equations \eqref{constrainteqns}. Similarly, the time component of
the momentum constraint, $H_t=0$, implies the incompressibility condition\footnote{Recall that we dropped the superscript
$(0)$ in \eqref{constrainteqns}.} $\nabla_i^{(0)}v^i=0$,
as does the Hamiltonian constraint, $H=0$. Finally, the $i$ component of
the momentum constraint,
$H_i=0$, implies the remaining equation given in \eqref{constrainteqns}.
In terms of the perturbation, the variables $v^i,w,p$ in \eqref{constrainteqns}
are given by
\begin{align}
\label{quantdef}
v_i&\equiv-\delta g_{ti}^{(0)}\,,\cr
w&\equiv\delta a_t^{(0)}\,,\cr
p&\equiv-\frac{4\pi T}{G^{(0)}}\left(\delta g_{rt}^{(0)}-h^{ij(0)}g_{ir}^{(0)}\delta g_{tj}^{(0)}\right)
-h^{ij(0)}\frac{\partial_i G^{(0)}}{G^{(0)}}\delta g_{tj}^{(0)}\,.
\end{align}
where $h^{ij(0)}$ is the inverse metric for $h_{ij}^{(0)}$ 

We emphasise that evaluating the constraints close to the horizon picks out just the subset of the perturbation involving
$\delta a_t^{(0)}$, $\delta g_{ti}^{(0)}$ and $\delta g_{rt}^{(0)}$, as well as the sources $E,\zeta$. Furthermore,
for a given set of sources the resulting generalised Stokes equations 
are a closed set of equations for $\delta a_t^{(0)}$, $\delta g_{ti}^{(0)}$ and $\delta g_{rt}^{(0)}$.

We also emphasise that the derivation of the Stokes equations we have just outlined only depends on the local behaviour of the metric and the perturbation near the horizon. In particular it only assumed that one can locally foliate the spacetime with surfaces of constant $r$ near the horizon. We will return to this point later.

\subsection{Electric and heat currents}\label{currentsec}

We define the bulk electric current density $J^a=(J^t,J^i)$, which is a function of
$(r,x^i)$, as
\begin{align}
J^a=\sqrt{-g}Z{F}^{ar}\,.
\end{align}
In the radial decomposition we have 
\begin{align}
J^a=\pi^a\,.
\end{align} 
The current density of the dual field theory is obtained by evaluating $J^a$ at the
AdS boundary. 
Defining
$J^i_{(0)} \equiv J^i|_{r=0}$ we find the expression for $J^i_{(0)}$
given in \eqref{expforJ}.

For the heat current, as in \cite{Donos:2015gia,Banks:2015wha} we first define:
\begin{align}\label{defgee}
G^{\mu\nu}\equiv-2\nabla^{[\mu}k^{\nu]}-\frac{2Z}{D-2}k^{[\mu}{F}^{\nu]\sigma}A_\sigma-\frac{1}{D-2}[(3-D)\theta+\varphi]Z{F}^{\mu\nu}\,,
\end{align}
where $\varphi\equiv i_kA$ and we write $i_kF\equiv \psi+d\theta$ for a globally defined function $\theta$. We
will take $\theta=-A_t$, so that $\varphi=-\theta=A_t$ and $\psi_\nu=\partial_t A_\nu$. 
We then define the bulk heat current density, $Q^i$, via
\begin{align}
Q^a\equiv\sqrt{-g}G^{ar}.
\end{align}
Using the radial decomposition we can express $Q^i$ as 
\begin{align}\label{qsd}
Q^i
&=-2{\pi^i}_t-\pi^iA_t\,.
\end{align}
An explicit calculation shows that this is time independent. Furthermore, $Q^i$ at the 
AdS boundary, $Q^i_\infty $, is the heat current density of the dual field theory\footnote{At the AdS boundary we have $2(\pi^i{}_t)_\infty=(\sqrt{-g}t^i{}_t)_\infty$ where $t$ is the stress tensor.}.
We also note that at the horizon, we have $Q^i_{(0)}=4\pi T\sqrt{h^{(0)}}v^i$. 

Notice that $G$ is not gauge independent. However, this does not affect our results.
We should only allow gauge transformations that leave $i_k A$ fixed
at the $AdS$ boundary, to ensure that we do not change the chemical potential, and also
at the horizon, to maintain regularity. We then see that such gauge transformations will
not change $Q^i$ at the horizon nor at the AdS boundary.

A subtlety, which we will return to below, is that the electric and heat currents at the boundary,
$J^i_\infty$ and $Q^i_\infty$, will be divergent in general and need to
be renormalised by boundary counterterm contributions.

\subsubsection{Radial evolution of currents}
We now examine the radial evolution\footnote{Recall that in this section we are assuming
that the radial coordinate is defined all the way from the AdS boundary down to the black hole. This condition
is relaxed in the next section.} of the bulk current densities $J^i,Q^i$. The equation of motion for the gauge field is:
\begin{align}\label{geom}
\partial_\mu(\sqrt{-g}Z{F}^{\mu\nu})&=0\,,
\end{align}
which implies the conditions
\begin{align}
\partial_rJ^i&=\partial_j(\sqrt{-g}Z{F}^{ji})+\sqrt{-g}Z{F}^{ij}\zeta_j,\cr
\partial_iJ^i&=J^i\zeta_i\,.\label{drJ}
\end{align}
We also have $\partial_t J^t=-J^i\zeta_i$; the local charge density is changing in time due to the presence
of the sources $E,\zeta$.

A little more work is required for the heat current. Using the equations of motion, $k^{\mu}\nabla_\mu \phi=0$ and the fact that
\begin{align}\label{kcons}
\nabla_\mu k^\mu=0,\qquad
\nabla_\mu\nabla^{(\mu}k^{\nu)}=(\tfrac{1}{2}\nabla_\mu \zeta^\mu)k^\nu-\tfrac{1}{2}(dk)^{\nu\rho}\zeta_\rho\,,
\end{align}
we can show that
\begin{align}
\nabla_\mu G^{\mu\nu}
&=\left(-\nabla_\mu \zeta^\mu+\frac{2V}{D-2} \right)k^\nu+(dk)^{\nu\rho}\zeta_\rho+\frac{D-3}{D-2}Z{F}^{\nu\mu}\psi_\mu-\frac{ZA_\sigma\mathcal{L}_k({F}^{\nu\sigma})}{D-2}\,,\label{Gder}
\end{align}
where $\psi$ was defined below \eqref{defgee}.
It is worth highlighting that unlike for the purely electric black hole solutions studied in \cite{Donos:2015gia,Banks:2015wha}, in general $(dk)^{\nu\rho}\zeta_\rho\not\propto k^\nu$.
From this expression we deduce
\begin{align}
\partial_rQ^i&=\partial_j(\sqrt{-g} G^{ji})+2\sqrt{-g}G^{ij}\zeta_j+\sqrt{-g}Z{F}^{ij} E_j\,,\cr
\partial_iQ^i&=2Q^i\zeta_i+J^iE_i.\label{drQ}
\end{align}

\subsubsection{Magnetisation Currents}
We begin by considering
the currents for the background black holes (i.e. when we set $E=\zeta=0$ and hence the whole of the
perturbation to zero). One can show that $J^i$ and $Q^i$ then both vanish at the horizon. Hence, upon integrating 
\eqref{drJ} and \eqref{drQ} we deduce that the holographic current densities for the background black holes, denoted with the superscript $(B)$,
are given by magnetisation current densities of the form
\begin{align}\label{jayback}
J^{(B)i}_\infty&=\partial_j M^{(B)ij}\,,\nn
Q^{(B)i}_\infty&=\partial_j M^{(B)ij}_T\,,
\end{align}
where the local magnetisation density, 
$M^{ij}(x)$, and the local thermal magnetisation density, 
$M^{ij}_T(x)$, are given by
\begin{align}\label{emmslocal}
M^{ij}&=-\int_0^\infty dr\sqrt{-g}Z{F}^{ij}\,,\cr
M_T^{ij}&=-\int_0^\infty dr\sqrt{-g}G^{ij}\,.
\end{align}
Clearly $\partial_i J^{(B)i}_\infty=\partial_i Q^{(B)i}_\infty=0$. 
If we now assume, for simplicity, that we have a periodic lattice with
$\Sigma_d$ (effectively) a torus, as in section \ref{usersguide}, then
we deduce that the current flux densities for the background black holes, 
defined as in \eqref{avecur}, \eqref{avecur2}, must vanish
$\bar{{J}}^{(B)i}_\infty=\bar{{Q}}^{(B)i}_\infty=0$.

When we switch on $E,\zeta$, we want to study the local ``transport current densities",
$\mathcal{J}^i(x)$, $\mathcal{Q}^i(x)$ of the boundary theory that are obtained by subtracting off magnetisation currents generated by $M$, $M_T$. 
Following the lead of \cite{PhysRevB.55.2344}, these can be defined at the holographic boundary via:
\begin{align}\label{caljdefs}
\mathcal{J}^i&\equiv J^i_{\infty}+M^{(B)ij}\zeta_j-\partial_j M^{ij},\cr
\mathcal{Q}^i&\equiv Q^i_{\infty}+M^{(B)ij}E_j+2M_T^{(B)ij}\zeta_j-\partial_j M_T^{ij}\,,
\end{align}
Notice that the terms on the right hand side added to $J^i_{\infty}$ and $Q^i_{\infty}$ ensure
\footnote{Notice that the terms $\partial_j M^{ij}$ and
$\partial_j M_T^{ij}$, which contain both background terms and terms linear in the perturbation, 
are not necessary to ensure this and furthermore they also make no contribution to the flux densities 
$\bar{\mathcal{J}}^i$ and $\bar{\mathcal{Q}}^i$, consistent with \eqref{initialbarj}.}
that 
\begin{align}
\partial_i \mathcal{J}^i=0\,,\qquad 
\partial_i\mathcal{Q}^i=0\,,
\end{align}
as one can deduce
from (\ref{drJ}), (\ref{drQ}) and \eqref{jayback}.

By integrating (\ref{drJ}), (\ref{drQ}) we now deduce that these local transport current densities are exactly the same
as the local current densities at the black hole horizon:
$\mathcal{J}^i(x)=J^i_{(0)}(x)$ and $\mathcal{Q}^i(x)=Q^i_{(0)}(x)$. Specialising now to the case of 
a periodic lattice with $\Sigma_d$ (effectively) a torus, then we deduce that the transport current flux densities, 
relevant for the DC conductivity, are given by the horizon current flux densities:
\begin{align}\label{reneq}
\bar{\mathcal{J}}^i=\bar J^i_{(0)},\qquad
\bar{\mathcal{Q}}^i=\bar Q^i_{(0)}\,,
\end{align}
where
\begin{align}\label{tfd2}
\bar{\mathcal{J}}^i&=\bar J^i_{\infty}+\overline{M^{{(B)}ij}\zeta_j},\cr
\bar{\mathcal{Q}}^i&=\bar Q^i_{\infty}+\overline{M^{(B)ij} E_j}+2\overline{M^{(B)ij}_{T}\zeta_j}\,.
\end{align}

We now write the closed one-form sources 
as $E=\bar E_idx^i+de$, $\zeta=\bar \zeta_idx^i+dz$, where $e,z$ are periodic functions and
$\bar E_i,\bar \zeta_i$ are constants. If we solve the generalised Stokes equations at the horizon we obtain the current densities $J^i_{(0)}, Q^i_{(0)}$
at the horizon as functions\footnote{Recall that we showed in section
\ref{genpropsstokes} that the currents at the horizon are independent of the
exact parts of $E$ and $\zeta$.} of $\bar E_i,\bar \zeta_i$. We can then obtain the constant current flux densities 
$\bar J^i_{(0)}, \bar Q^i_{(0)}$ by integrating
as in \eqref{avecur}, 
\eqref{avecur2}, and hence obtain,  via \eqref{reneq}, $\bar{\mathcal{J}}^i,\bar{\mathcal{Q}}^i$
as functions of the $\bar E_i,\bar \zeta_i$ and hence the thermoelectric DC conductivity via \eqref{bigform2}.

In appendix \ref{magcom} we make some additional comments concerning the total magnetisation
of the equilibrium black holes using notation that we introduce in the next section.

To conclude this section we return to the issue of counterterms.
We noted above that $J^i_\infty$ and $Q^i_\infty$ will be divergent and receive contributions
from boundary counterterms. Similarly $M^{ij}$ and $M^{ij}_T$ defined in 
\eqref{emmslocal} and appearing in \eqref{caljdefs} will also need to be renormalised. We explain in
appendix \ref{holren} how, in effect, we can replace the quantities on the right hand side with renormalised quantities leaving
$\mathcal{J}^i(x)$ and $\mathcal{Q}^i(x)$ unchanged. This cancellation of divergences is consistent with the fact
that we have already shown that $\mathcal{J}^i(x)$ and $\mathcal{Q}^i(x)$ are finite quantities through the relations 
$\mathcal{J}^i(x)=J^i_{(0)}(x)$ and $\mathcal{Q}^i(x)=Q^i_{(0)}(x)$. 
Similar comments apply to the fluxes $\bar{\mathcal{J}}^i$ and $\bar{\mathcal{Q}}^i$.

\section{Derivation for general black hole solutions}\label{der2}
In this section we will generalise the analysis of the previous section to cover black hole solutions discussed in section \ref{sec2},
for which we do not assume that there is a single black hole horizon with the coordinates $(t, r,x^i)$ globally defined everywhere outside
the horizon. 
This covers the possibility that there are multiple black hole horizons all
with the same temperature. We do assume, however, that there is
a time coordinate, $t$, globally defined outside the event horizons,
with the Killing vector given by $k=\partial_t$. At the AdS boundary the solution behaves as
in \eqref{asmet} with spatial sections $\Sigma_d$. We do not assume that
the topology of the black hole horizons are also $\Sigma_d$.

It is convenient to write the black hole solution combined with a suitable linear perturbation in the following compact form
\begin{align}\label{gfan}
ds^2&=-H^2 (dt+\alpha +t \zeta)^2+ds^2(M_{D-1})\,,\nn
A&=a_t(dt+\alpha+t\zeta)-tE +\beta
\end{align}
Here $H, a_t$ and the scalar field $\phi$ are functions on $M_{D-1}$, and the metric $ds^2(M_{D-1})$ is also
independent of $t$. 
In addition $\alpha,\beta$ are, locally, one-forms on $M_{D-1}$: it is important that we allow $\alpha,\beta$
to be not globally defined in order that $d\alpha$ and $d\beta$ can define non-trivial cohomology classes, corresponding
to NUT and magnetic charges, respectively.
We emphasise that all of these quantities include the background black hole solution as well as 
a linearised perturbation. 
In addition $E,\zeta$ are one-forms on 
$M_{D-1}$ and we demand that they are closed: $d\zeta=dE=0$. 
Locally one can remove the explicit time dependence from this ansatz\footnote{This can be seen by writing, locally, $E=de, \zeta=dz$ and then doing the change of coordinate $t=(1-z)\tilde t$ as well as the gauge transformation $A\to A+d(\tilde t e)$.} but not globally. 
At the AdS boundary we assume that the perturbation is chosen so that the only source terms are given by $E,\zeta$. In the local coordinates
$(t,r,x^i)$ near the boundary, used in \eqref{asmet}, we have that $E,\zeta$ approach closed one-forms on $\Sigma_d$. 

\subsection{Stokes equations}
Near each black hole horizon we can introduce local coordinate systems $(t,r, x^i)$ and then carry out a local radial decomposition
of the equations of motion. Evaluating the constraint equations near each horizon, exactly as outlined in section \ref{conhor}, we will obtain a generalised set of Stokes equations on each horizon. Note that
if a given horizon has topology $\tilde\Sigma_d$, then the source terms will approach closed one-forms on $\tilde\Sigma_d$, with their precise form
fixed by solving the closure condition in the bulk $dE=d\zeta=0$,
as well as by the specific closed forms they approach on $\Sigma_d$ at the AdS boundary.

\subsection{Electric and heat currents}\label{currentsmarktwo}
We now consider the electric and magnetic currents, working at linearised order in the perturbation. Using the same notation, the one-form dual to the vector $k=\partial_t$ has the form $k=-H^2 (dt+\alpha +t \zeta)$. We also have
\begin{align}\label{kcons}
\nabla_\mu k^\mu=0,\qquad
\nabla_\mu\nabla^{(\mu}k^{\nu)}=(\tfrac{1}{2}\nabla_\mu \zeta^\mu)k^\nu-\tfrac{1}{2}(dk)^{\nu\rho}\zeta_\rho\,,
\end{align}
and we note that the last term in \eqref{kcons} is not proportional to $k$ when $d\alpha\ne 0$ in the background solution.

For the electric current we first note that $F=dA$ with
\begin{align}\label{fexp}
F&=-H^{-2}(da_t-a_t\zeta+E)\wedge k+a_td\alpha+d\beta+\alpha\wedge(E-a_t\zeta)\,,
\end{align}
and we will shortly use the fact that the gauge equation of motion is $d*Z(\phi)F=0$.
For the heat current we consider the two-form
\begin{align}\label{geeform}
G=-dk
-\frac{Z(\phi)}{D-2} k\wedge s-\frac{1}{D-2}\,\left[\left(3-D\right)\,\theta+\varphi \right]Z(\phi)\,F\,,
\end{align}
where we have defined $s\equiv -i_AF$,
$\varphi=i_{k}A$ and $i_{k}F=d\theta +\psi$, with $\psi$ a one-form and $\theta$ a globally defined function.
For our set up we take $\varphi=-\theta=a_t$ and $\psi=-E+a_{t}\,\zeta$.
We can show that the equations of motion imply
\begin{align}\label{effgeeeqn}
d*G=&
(-1)^D\left(-\nabla_\mu \zeta^\mu+\frac{2V}{D-2} \right)*k
-\zeta\wedge * dk\nn
&+\frac{3-D}{D-2}Z(\phi)\psi\wedge * F
+\frac{1}{D-2}A\wedge L_k*(Z(\phi)F)\,.
\end{align}

We proceed by calculating $d i_k *Z(\phi)F=L_k*Z(\phi)F=Z(\phi)\partial_t*F$, where the first equality used the gauge equation
of motion, and similarly for $di_{k}*G$. 
Proceeding in this way (and using \eqref{stareff})
we obtain the key results
\begin{align}\label{dstar}
d i_k *(Z(\phi)F)&=-\zeta\wedge m\,,\nn
di_{k}*G&= -E\wedge m-2\zeta\wedge m_T\,,
\end{align}
where the $(D-3)$-forms $m$ and $m_T$ are given by 
the following expressions for the background black holes
\begin{align}\label{defemms}
m=-H Z\bar *(a_td\alpha+d\beta)\,,\qquad
m_T=-H^3\bar *d\alpha -a_tm\,,
\end{align}
and $\bar *$ refers to the metric $ds^2(M_{D-1})$. Note that \eqref{dstar} generalises (\ref{drJ}), (\ref{drQ}) of the previous section. 
Since $E,\zeta$ are closed we can deduce from \eqref{dstar} that $m,m_T$ are also closed:
\begin{align}
dm=0\,,\qquad dm_T=0\,.
\end{align}

\subsection{DC conductivity}
At the AdS boundary of the background solutions, at fixed $t$, we introduce a basis, $C_a$, $a=1,\dots, b_{d-1}(\Sigma_d)$ of $d-1$ closed cycles. We can then define the holographic current fluxes through these cycles via
\begin{align}\label{defjay}
\bar J^a_\infty\equiv - \int_{C_a}i_{k}*Z(\phi)F\,,
\qquad
\bar Q^a_\infty\equiv - \int_{C_a}i_{k}*G\,.
\end{align}
We emphasise that $\bar J_\infty^a$, $\bar Q_\infty^a$ depend on the choice of
cycles $C_a$ and not just their homology class. 
This is because $di_k *Z(\phi)F|_\infty=-\zeta\wedge m|_\infty\ne0$, in general.
This is a manifestation of the fact that these are not
the correct transport current fluxes due to the magnetisation of the background, parametrised
by $m$. We remedy this as follows.

Consider a $d$-dimensional surface $S_a$ 
in the bulk spacetime which has boundary $C_a$ at the $AdS$ boundary and possible additional boundaries
$C^a_{H_i}$ at the black hole horizons. We first define the horizon current fluxes via
\begin{align}
\bar J^a_{H_i}&\equiv - \int_{C^a_{H_i}}i_{k}*Z(\phi)F\,.
\end{align}
Since $di_k *Z(\phi)F|_{H_i}=-\zeta\wedge m|_{H_i}=0$ we deduce that $\bar J^a_{H_i}$ only depends
on the homology class of $C^a_{H_i}$. We next define the transport current fluxes via
\begin{align}
\bar{\mathcal{J}}^a
&\equiv\bar J^a_\infty +\int_{S_a}  \zeta\wedge m\,.
\end{align}
If we continuously deform $S_a$ in the bulk, keeping
$C_a$ and $C^a_{H_i}$ fixed,  
then $\bar{\mathcal{J}}^a$ does not change since $d(\zeta\wedge m)=0$. Furthermore
while each of the two terms in $\bar{\mathcal{J}}^a$ does depend on the particular representative chosen
for $C_a$, the sum only depends on the homology class of $C_a$.
With these definitions in hand, after integrating \eqref{dstar} along $S_a$ we immediately deduce that
\begin{align}
\bar{\mathcal{J}}^a_\infty&=\sum_i \bar J^a_{H_i}\,.
\end{align}

Similarly defining
\begin{align}
\bar{\mathcal{Q}}^a_\infty
&\equiv\bar Q^a_\infty +\int_{S_a}(E\wedge m+\zeta\wedge 2m_T)\,,\nn
\bar Q^a_{H_i}&\equiv - \int_{C^a_{H_i}}i_{k}*Z(\phi)F\,,
\end{align}
we have
\begin{align}
\bar{\mathcal{Q}}^a_\infty&=\sum_i \bar Q^a_{H_i}\,.
\end{align}

We have shown that transport current flux densities 
$\bar{\mathcal{J}}^a_\infty$, $\bar{\mathcal{Q}}^a_\infty$, 
are equal to the sum of the  
$\bar J^a_{H_i}$, $\bar Q^a_{H_i}$ at the horizons. In turn the latter are fixed by solving the generalised Stokes equations in terms of
the cohomology class of the sources $(E,\zeta)$ at the horizon (since the exact parts do not affect $\bar J^a_{H_i}$, $\bar Q^a_{H_i}$).
In turn the cohomology classes of the sources at the horizon are fixed by the cohomology classes of the sources at the AdS boundary.
By expanding $E=\bar E^a\eta^a_\infty+\dots$, $E=\bar E^a\eta^a_\infty+\dots$ at the AdS boundary we can extract the DC conductivity.

We can now make the connection with the formulae in the last section explicit. We first note that when $\Sigma_d$ is closed
it is helpful to use Poincar\'e duality. The cycle $S_a$, with fixed homology classes $C_a$ and $C_{H_i}^a$
is Poincare dual to a closed one-form $\eta^a$ on $M_{D-1}$ which approaches closed
one-forms $\eta^a_\infty$ on $\Sigma_d$ at the AdS boundary and $\eta^a_{H_i}$
at the horizons. We then have
\begin{align}\label{pdexps}
\bar{\mathcal{J}}^a
&=-\int_{\Sigma_d}\eta_\infty^a\wedge i_{k}*Z(\phi)F  +\int_{M_{D-1}}{\eta^a\wedge\zeta\wedge m}\,,\nn
\bar{\mathcal{Q}}^a&= -\int_{\Sigma_d}\eta_\infty^a\wedge i_{k}*G  +\int_{M_{D-1}}{\eta^a\wedge (E\wedge m+\zeta\wedge 2m_T)}\,,
\end{align}
and
\begin{align}
\bar{{J}}^a_{H_i}
&=-\int_{H_i}\eta_{H_i}^a\wedge i_{k}*Z(\phi)F \,,\nn
\bar Q^a_{H_i}&= -\int_{H_i}\eta_{H_i}^a\wedge i_{k}*G\,.
\end{align}
These definitions only depend on the cohomology class of $\eta^a$, $\eta^a_\infty$ and $\eta^a_{H_i}$.

We now consider black holes with a single black hole horizon, with $\Sigma_d=T^d$ and for which we
can choose the coordinates $(r,x^i)$ globally outside the black hole horizon. We can then take a basis
of closed one-forms to be $\eta^i=\eta^i_\infty=\eta^i_{H}=(\prod_i L_i)^{-1}\{dx^i\}$ (which in general are not harmonic). 
After substituting into \eqref{pdexps}, after a little calculation we obtain the transport current flux densities \eqref{tfd2}.

In appendix \ref{magcom} we make some additional comments concerning $m$, $m_T$ and the total magnetisation
of the equilibrium black holes as calculated from variations of the free energy.

\section{Examples}\label{exsect}
In this section we will solve the Stokes equations for several examples and hence
obtain explicit expressions for the DC conductivity using the procedure 
outlined in section \ref{usersguide}. We will only consider solutions in which the
coordinates $(t,r,x^i)$ are globally defined outside a single black hole horizon and
moreover just consider periodic lattices, for which $\Sigma_d$ can be taken to be a torus.

We will express some of our results in terms of the total averaged charge density, $\rho$, of the dual CFT. 
We recall that the time component of the bulk electric current density has the form $J^t=\sqrt{-g}Z(\phi)F^{tr}$. At the black hole horizon we thus have $\rho_H\equiv J^t|_H=\sqrt{-g_0}Z^{(0)}A^{(0)}_{t}$, as in \eqref{Hquants}.
Recalling the equation of motion for the gauge field is $d*[Z(\phi)F]=0$, we have
\begin{align}\label{rhotot}
\rho\equiv \frac{1}{\text{vol}_d}\int d^dx (\sqrt{-g}Z (\phi)F^{tr})|_\infty
=\frac{1}{\text{vol}_d}\int d^dx \rho_H\,,
\end{align}
where $\text{vol}_d\equiv \int d^dx \sqrt{-h^{(\infty)}}$ is the volume of the
spatial metric at the $AdS$ boundary.

\subsection{Extra Scalars and Q-lattices}\label{qlattice}

If we replace our bulk Lagrangian with one containing several scalars, $\phi^I$, 
with the functions $V,Z$ depending on all of the scalars and the kinetic-energy terms generalised via
\begin{align}
-\frac{1}{2}\partial\phi^2\to -\frac{1}{2}{\cal G}_{IJ}(\phi)\partial\phi^I\partial\phi^J\,,
\end{align}
then this leads to the Stokes equations \eqref{constrainteqns} as before, with the only change given by
\begin{align}
\nabla_j\phi^{(0)}\nabla_i\phi^{(0)}v^{j}
\to
{\cal G}_{IJ}(\phi^{(0)})\nabla_j\phi^{I(0)}\nabla_i\phi^{J(0)}v^{j}\,.
\end{align}
We can now obtain the DC conductivities for general 
Q-lattices in a magnetic field, generalising the results of
\cite{Blake:2015ina} and \cite{Banks:2015wha}. The derivation is very similar to 
 \cite{Banks:2015wha} so we shall be brief.

In order to obtain the Q-lattice solutions we assume that the model admits $n$ shift symmetries of the scalars:
\begin{align}
\phi^{I_{\alpha}}\rightarrow \phi^{I_{\alpha}}+\epsilon^{I_{\alpha}}\,,
\end{align}
with $\alpha=1,\dots,n$ and constant $\epsilon^{I_{\alpha}}$. These could be, for example, shifts of the phase of a complex scalar.
Notice that the function $Z$ must be independent of the scalars $\phi^{I_{\alpha}}$.
Using the coordinates $x^\mu=(t,r,x^i)$, the black hole solutions are constructed using an ansatz in which 
the scalars associated with these shift symmetries take the form
\begin{align}
\phi^{I_{\alpha}}=\mathcal{C}^{I_{\alpha}}{}_{j}\,x^{j}\,,
\end{align}
everywhere in bulk with $\mathcal{C}$ a constant $n$ by $d$ matrix. For simplicity 
of presentation we assume that all spatial coordinates are involved. 
We then take the remaining scalar fields and the metric\footnote{Note that this implies
in particular that we have set possible NUT charges to zero. It is simple to include them e.g. see appendix \ref{nutchge}.}  to only depend on the radial coordinate. Since $Z$ is independent of the scalars at the horizon we have $Z^{(0)}$ is a constant.
For the gauge field we take 
$A = a_t dt+a_i dx^i + F_{ij}x^idx^j$ with $a_t$, $a_i$ functions of $r$ and 
{\it constant} magnetic field $F_{ij}$.
The spatial metric at the AdS boundary is taken to be flat and hence the
spatial metric on the horizon, $h_{ij}^{(0)}$, is also flat, but there can be a change of length scales
with respect to the coordinates.

In perturbing these black hole solutions, the sources are taken to be $E=\bar E_idx^i$, $\zeta=\bar \zeta_idx^i$ with $\bar E_i,\bar \zeta_i$ constants. All background quantities at the horizon entering the Stokes equations (\ref{constrainteqns}) are constant and hence they are solved with constant $v^{i}$, $p$ and $w$. 
The fluid velocity is given by
\begin{align}
v^{i}&=\,\left(\mathcal{M}^{-1} \right)^{ij}\,\left(4\pi T\zeta_{j}+\frac{4\pi \rho}{s}\mathcal{N}_{j}{}^kE_{k}\right)\,,
\end{align}
where we have defined the constant $d\times d$ matrices 
\begin{align}
\mathcal{M}_{ij}=\mathcal{D}_{ij}-\frac{4\pi \rho}{s}{F}_{ij}+Z^{(0)}{F}_{ik}{F}{}_j{}^k\,,\qquad
\label{Mijdefn}
\mathcal{N}_i{}^j=\delta_i^j+\frac{sZ^{(0)}}{4\pi \rho}{F}{}_{i}{}^j\,,
\end{align}
and the symmetric $d\times d$ matrix:
\begin{align}
\mathcal{D}_{ij}&=G_{I_{\alpha_{1}} I_{\alpha_{2}}}\,\mathcal{C}^{I_{\alpha_{1}}}{}_{i}\,\mathcal{C}^{I_{\alpha_{2}}}{}_{j}\,.
\end{align}
Furthermore, the averaged charge density, $\rho$, defined in 
\eqref{rhotot}
and the entropy density, $s$, are given by 
\begin{align}\label{rhoands}
\rho=\rho_H&=\sqrt{h^{(0)}}{Z^{(0)}A_{t}^{(0)}},\qquad s=s_H=4\pi \sqrt{h^{(0)}}\,,
\end{align}
both of which are constants.

The current densities $J^i_{(0)},Q^i_{(0)}$ at the
horizon can easily be obtained and are constant. From \eqref{bigform2} we can then deduce
\begin{align}
\sigma^{ij}&=\frac{s\,Z^{(0)}}{4\pi}{h^{ij}}^{(0)}+ 
\frac{4\pi \rho^2}{s}\mathcal{N}^{i}{}_k\left(\mathcal{M}^{-1} \right)^{kl}\mathcal{N}_l{}^{j}\,,\cr
\alpha^{ij}&=4\pi\rho \, \mathcal{N}^{i}{}_k\left(\mathcal{M}^{-1} \right)^{kj}\,,\cr
\bar\alpha^{ij}&=4\pi\rho \, \left(\mathcal{M}^{-1} \right)^{ik}\mathcal{N}_k{}^j ,\cr
\bar{\kappa}^{ij}&=4\pi T s\,\left( \mathcal{M}^{-1}\right)^{ij}\,.\label{Qlattcond}
\end{align}
We easily see that we have the Onsager relations
$\alpha^T ({F})=\bar\alpha(-{F})$, $\bar\kappa^T({F})=\bar\kappa(-{F})$
and $\sigma^T({F})=\sigma(-{F})$.
Note that in general we do not have $\alpha=\bar\alpha$.
Also, the conductivity when $Q=0$, $\sigma_{Q=0}\equiv  \sigma-T\alpha\bar\kappa^{-1}\bar\alpha$, is given by
\begin{align}\label{sqzex}
\sigma^{ij}_{Q=0}=\frac{s\,Z^{(0)}}{4\pi}{h^{ij}}^{(0)}\,.
\end{align}
Observe for this case, and recalling  \eqref{Hquants},
that we have $\sigma^{ij}_{Q=0}=\Sigma^{ij}_H$.

\subsection{Comparison with \cite{Blake:2015ina}}\label{comparsec}

We apply the above formulae to the anisotropic Q-lattice black holes with magnetic fields in $D=4$  considered in
\cite{Blake:2015ina} (see also \cite{Kim:2015wba}). In particular, we can consider the action
\begin{align}
S = \int \mathrm{d}^4x \sqrt{-g} \bigg [ R - \frac{1}{2}[( \partial \phi)^2 + \Phi_1(\phi) (\partial {\chi_1})^2 + \Phi_2(\phi) (\partial {\chi_2})^2] + V_T(\phi) - \frac{Z(\phi)}{4} F^2 \bigg]\,,
\label{anisotropic}
\end{align}
where we now break translational invariance in the $x$ and $y$ directions by constructing background solutions with $\chi_1 = k_1 x$ and $\chi_2 = k_2 y$. The metric is of the form
\begin{align}
ds^2 = - U dt^2 + U^{-1} dr^2 + e^{2V_1}dx^2 + e^{2V_2} dy^2\,,
\end{align}
and $F_{xy}=B$ is a constant. Dropping the superscript on the constant $Z^{(0)}$ for simplicity
of presentation, we then find
\begin{align}
\sigma^{ij}
&=\frac{1}{ \Delta}\left(\begin{array}{cc} e^{V_2+V_1}k_2^2 \Phi_2(\rho^2+Ze^{2V_2}k_1^2\Phi_1 +B^2Z^2) & B\rho(\rho^2+Z e^{2V_2}k_1^2\Phi_1  +Ze^{2V_1}k_2^2\Phi_2 +B^2Z^2)\\ -B\rho(\rho^2+Z e^{2V_1}k_2^2\Phi_2  +Ze^{2V_2}k_1^2\Phi_1 +B^2Z^2) &e^{V_1+V_2}k_1^2\Phi_1 (\rho^2+Ze^{2V_1} k_2^2 \Phi_2+B^2Z^2)\end{array}\right)\,,\cr
\cr
\alpha^{ij}&= \frac{s}{ \Delta}\left(\begin{array}{cc} \rho k_2^2\Phi_2e^{V_1+V_2}  & B(\rho^2  + Ze^{2V_2}k_1^2\Phi_1 +B^2Z^2)\\ -B(\rho^2   +Ze^{2V_1}k_2^2\Phi_2 +B^2Z^2)&\rho k_1^2\Phi_1e^{V_1+V_2} \end{array}\right)\,,\cr
\cr
\bar\alpha^{ij}&=\, \frac{s}{ \Delta}\left(\begin{array}{cc} \rho k_2^2\Phi_2e^{V_2+V_1} & B(\rho^2  +Ze^{2V_1}k_2^2\Phi_2 +B^2Z^2)\\ -B(\rho^2  +Ze^{2V_2}k_1^2\Phi_1 +B^2Z^2) &\rho k_1^2\Phi_1e^{V_2+V_1}\end{array}\right),\cr
\cr
\bar{\kappa}^{ij}&=\,\frac{sT}{ \Delta}\left(\begin{array}{cc} 4\pi  e^{2V_2}(e^{2V_1}k_2^2\Phi_2 +B^2Z) & s \rho B\\ -s \rho B&4\pi  e^{2V_1}(e^{2V_2}k_1^2\Phi_1 +B^2Z)\end{array}\right)\,,\label{aristoscond}
\end{align}
where $\Delta\equiv B^2\rho^2+(e^{2V_2}k_1^2\Phi_1 +B^2Z)(e^{2V_1} k_2^2 \Phi_2+B^2Z)$ and the right hand side is evaluated at $r=0$. Furthermore,
$\rho,s$ are defined as in \eqref{rhoands}.
The expressions agree with \cite{Blake:2015ina} and we note that in general $\alpha^{xy}$ is not equal to $\bar{\alpha}^{xy}$.

The expression for $\sigma_{Q=0}$ given in \eqref{sqzex} is a simple diagonal matrix. We can also calculate the thermal
conductivity for zero current flow, $\kappa^{ij}\equiv\bar{\kappa}^{ij}-T\bar{\alpha}^{ik}\sigma^{-1}_{kl}\alpha^{lj}$. 
Interestingly it can be obtained by making the substitution $B\to -\rho$, $\rho\to B$ and $Z\to 1/Z$ in the expression for $\bar\kappa$.
The reason for this will be clarified in appendix \ref{sduality}.

\subsection{Homogeneous black holes}
We can also apply the results we have obtained to the special case of homogeneous black holes in the absence
of a lattice. We simply take the conductivity results for the Q-lattices given in (\ref{Qlattcond}) and set $\mathcal{D}_{ij}=0$. In fact, we can simplify (\ref{Qlattcond}) further in this case, and we obtain
\begin{align}\label{homogexp}
\sigma^{ij}&=-\rho({F}^{-1} )^{ij}\,,\qquad
\alpha^{ij}=\bar\alpha^{ij}=-s \, ({F}^{-1})^{ij}\,,\nn
\bar{\kappa}^{ij}&=- \frac{T s^2}{\rho}\,(\mathcal{N}^{-1})^i{}_k({F}^{-1})^{kj}\,.
\end{align}
Note that we have $\alpha^{ij}=\bar\alpha^{ij}=\frac{s}{\rho}\sigma_{ij}$. Furthermore, we see that we require ${F}_{ij}$ to be invertible as a matrix if we want the DC conductivities to be finite, which requires $d$ to be even. 

A particular example is the standard dyonic black hole solution in $D=4$ with $\phi=0$. It is convenient to define $Z(0)=1/e^2$. The solution is given by
\begin{align}\label{adsexact}
ds^2&=-Udt^2+U^{-1}d{r}^2+{r}^2(dx^2+dy^2),\cr
A&=A_t dt+\tfrac{1}{2}B(xdy-ydx),
\end{align}
with constant magnetic field $B$, $U=r^2-\frac{M}{{r}}+\frac{Q^2+H^2}{{r}^2}$ and $A_t=\mu(1-\frac{r_+}{{r}})$, with $M=r_+^{3} + \frac{Q^2+H^2}{r_+}$, $H=\frac{B}{2e}$ and $Q=\frac{\mu r_+}{2e}$. The event horizon is located at $r=r_+$ in these coordinates. The black hole has temperature $T=\frac{3r_+}{4\pi}(1-\frac{Q^2+H^2}{3r_+^4})$, 
charge density $\rho=2Q/e$, energy density $\epsilon=2M$ and entropy density $s=4\pi r_+^2$.
The conductivities for this black hole can be obtained from (\ref{homogexp}) (or alternatively directly from (\ref{aristoscond})): 
\begin{align}\label{resultadsrn}
\sigma^{ij}&=\frac{\rho}{ B}\left(\begin{array}{cc} 0 & 1 \\ -1 &0\end{array}\right)\,,\qquad\alpha^{ij}= \bar\alpha^{ij}=\frac{s}{ B}\left(\begin{array}{cc} 0 & 1\\ -1&0 \end{array}\right)\,,\cr
\bar{\kappa}^{ij}&=\,\frac{s^2Te^2}{B(\rho^2e^4+B^2)}\left(\begin{array}{cc}B&  \rho e^2 \\ - \rho e^2 &B\end{array}\right)\,.
\end{align}
The result for $\sigma$ matches that found in \cite{Hartnoll:2007ai} while the expressions for $\alpha,\bar\alpha,\bar\kappa$ are new.

The broad structure of these results have a simple heuristic interpretation. 
Suppose $\rho=0$. Consider applying a thermal gradient in the $x_1$ direction: $\zeta=(\zeta_1,0)$. Let us now
imagine the system to be made of charged ``particles" and ``holes" with opposite charge. The thermal
gradient will cause the particles and holes to move in the $x^1$ direction and the magnetic field will deflect them
in opposite directions in the $x^2$ direction. This sets up a net current in the $x^2$ direction, corresponding
to $\alpha^{12}\ne 0$, but there is no net momentum flow in this direction and so $\bar\kappa^{12}=0$. When $\rho\ne 0$
the charge density gives rise to $\bar\kappa^{12}\ne 0$ as well as $\sigma^{12}\ne 0$.

Other work on thermoelectric conductivity for homogeneous black holes appears in \cite{Hartnoll:2007ip,Lindgren:2015lia}.

\subsection{One-Dimensional Lattices}\label{onedtext}
We now shift gears and consider  inhomogeneous black hole solutions which only depend, periodically,
on one of the spatial coordinates, $x=x+L_1$. To illustrate
we just consider the $D=4$ case. On the horizon we assume the background black hole solutions
have the form\footnote{We demand that the ansatz for the full solution is
invariant under $y\to-y$, $t\to-t$, with the gauge-field going to minus itself
(which is a symmetry of the equations of motion).} 
\begin{align}\label{2onedform}
h^{(0)}_{ij}\,dx^{i}dx^{j}&=\gamma(x) \,dx^{2}+\lambda(x)dy^2\,,\nn
{F}^{(0)}_{xy}&\equiv B_H(x)\,,\nn
-(4\pi T){\chi}_y^{(0)}&\equiv \chi(x)\,,
\end{align}
and
${\chi}^{(0)}_x=0$. The scalar field at the horizon is a function of $x$, $\phi^{(0)}=\phi^{(0)}(x)$ and hence so
is $Z^{(0)}$.
We continue to take the electric and heat sources $E_i,\zeta_i$ to be constants.

The incompressibility condition $\partial_i(\sqrt{h^{(0)}}v^i)=0$ is solved by taking
\begin{align}
v^x=(\gamma\lambda)^{-1/2}\,v_{0}\,,
\end{align}
with $v_{0}$ a constant. 
The components of the current density at the horizon, \eqref{expforJ}, can be written
\begin{align}
{J^x_{(0)}}&=\frac{\lambda^{1/2}Z^{(0)}}{\gamma^{1/2}}(E_x+\partial_xw+\frac{A_t^{(0)}\gamma^{1/2}}{\lambda^{1/2}}v_{0}+B_Hv^y)\,,\label{2jxeqn}\\
{J^y_{(0)}}&=\frac{\gamma^{1/2}Z^{(0)}}{\lambda^{1/2}}(E_y+\lambda A_t^{(0)}v^y-\frac{B_H}{(\gamma\lambda)^{1/2}}v_{0})\,,\label{2jyeqn}
\end{align}
The equation $\partial_i {J^i}^{(0)}=0$ implies that ${J^{x}}^{(0)}$ is a constant. 

With a little effort we can write the $x$ component of the final Stokes equation in the form
\begin{align}\label{2navx}
&2\,v_{0}\,\partial_{x}\left( \gamma^{-1/2}\,\partial_{x}\lambda^{-1/2}\right)-Y\,v_{0}+\frac{\gamma^{1/2}A_{t}^{(0)}}{\lambda^{1/2}}J^x_{(0)}-\partial_{x}p\cr
&\qquad\qquad+\left(\partial_x\chi -Z^{(0)} A_t^{(0)} B_H\right)v^y+\frac{B_H}{(\gamma \lambda)^{1/2}}J^y_{(0)}=-4\pi T\,\zeta_{x}\,,
\end{align}
where we have defined
\begin{align}
Y\equiv\frac{1}{\lambda^{5/2}\gamma^{1/2}}\left(\partial_{x} \lambda\right)^{2}
+\frac{1}{(\gamma\,\lambda)^{1/2}}\,\left(\partial_{x}\phi^{(0)}\right)^{2}+\frac{\gamma^{1/2} Z^{(0)}A_{t}^{(0)}{}^{2}}{\lambda^{1/2}}\,.
\end{align}
On the other hand the $y$ component has the form
\begin{align}\label{2keyeq}
\partial_x\left(\frac{\lambda^{3/2}}{\gamma^{1/2}}\partial_x v^y\right)
=\partial_x\chi  v_0+B_HJ^x_{(0)}-Z^{(0)} A_t^{(0)}(\gamma \lambda)^{1/2}E_y-(4\pi T)(\gamma \lambda)^{1/2}\zeta_y\,.
\end{align}

We need to use the above equations to solve for four unknowns, namely, the two constants $v_0$ and $J^x_{(0)}$ and the two functions $v^y$ and $J^y_{(0)}$ in terms of the sources. From this we can extract all of the thermoelectric conductivities. 
We have presented a few details of the calculation in appendix \ref{onedapp}
and here we just state the final result.

From the $x$-dependent
horizon quantities 
$\rho_H=(\gamma\lambda)^{1/2} Z^{(0)} A_t^{(0)}$, $s_H= 4\pi(\gamma\lambda)^{1/2}$ and $B_H$, we first define
the total charge density, $\rho$, the total entropy density, $s$ and the average magnetic charge $B$, via
\begin{align}\label{rsb}
{\rho}=\int \rho_H\,,\qquad 
{s}=\int s_H\,,\qquad
{B}=\int B_H\,,
\end{align}
where $\int\equiv L_1^{-1}\int_0^{L_1}dx$ is the period average. 
We emphasise that $\rho,s,B$ are constants.
We next define two periodic functions, $w_i$, given by
\begin{align}\label{defws}
w_1(x)=\frac{{\rho}}{B}\int^xB_H-\int^x \rho_H\, ,\qquad
w_2(x)=\frac{T{s}}{B}\int^xB_H- T\int^x s_H\, ,
\end{align}
where $\int^x\equiv \int^x_0 dx$.
From the $w_i$ and $\chi $ we then define the following periodic functions
\begin{align}\label{defus}
u_1(x)&=\int^x\frac{\gamma^{1/2}\chi }{\lambda^{3/2}}-\frac{\int\frac{\gamma^{1/2}\chi }{\lambda^{3/2}}}{\int\frac{\gamma^{1/2}}{\lambda^{3/2}}}\int^x\frac{\gamma^{1/2}}{\lambda^{3/2}}\,,\cr
u_2(x)&=\int^x\frac{\gamma^{1/2}w_1}{\lambda^{3/2}}-\frac{\int\frac{\gamma^{1/2}w_1}{\lambda^{3/2}}}{\int\frac{\gamma^{1/2}}{\lambda^{3/2}}}\int^x\frac{\gamma^{1/2}}{\lambda^{3/2}}\,,\cr
u_3(x)&=\int^x\frac{\gamma^{1/2}w_2}{\lambda^{3/2}}-\frac{\int\frac{\gamma^{1/2}w_2}{\lambda^{3/2}}}{\int\frac{\gamma^{1/2}}{\lambda^{3/2}}}\int^x\frac{\gamma^{1/2}}{\lambda^{3/2}}\,.
\end{align}
Finally we define the constant matrix
\begin{align}
\mathcal{U}_{ij}\equiv\int\frac{\lambda^{3/2}}{\gamma^{1/2}}\partial_xu_i\partial_xu_j\,,
\end{align}
which can simplified a little after substituting in \eqref{defus}.

In terms of these quantities, the components of $\sigma$ can expressed as:
\begin{align}
\sigma^{xx}&=0\,,\cr
\sigma^{yy}&=\mathcal{U}_{22}+\int\frac{\gamma^{1/2}Z^{(0)}}{\lambda^{1/2}}+\frac{{\rho}^2}{{B}^2}\int\frac{\gamma^{1/2}}{\lambda^{1/2}Z^{(0)}}-\frac{1}{X}\left(\mathcal{U}_{12}-\frac{{\rho}}{{B}}\int\frac{\rho_H}{\lambda Z^{(0)}}-\int\frac{B_HZ^{(0)}}{\lambda}\right)^2\,,\cr
\sigma^{xy}&=-\sigma^{yx}=\frac{\rho}{{B}}\,,\label{sig1d}
\end{align}
where $X=\int Y+\int\frac{B_H^2Z^{(0)}}{\gamma^{1/2}\lambda^{3/2}}+\mathcal{U}_{11}$.
The components of $\alpha$ and $\bar\alpha$ are given by:
\begin{align}
\alpha^{xx}&=\bar{\alpha}^{xx}=0\,,\cr
\alpha^{yy}&=\bar{\alpha}^{yy}=\frac{1}{T}\mathcal{U}_{23}+\frac{{s}{\rho}}{{B}^2}\int\frac{\gamma^{1/2}}{\lambda^{1/2}Z^{(0)}}\cr
&\qquad\qquad-\frac{1}{X}\left(\mathcal{U}_{12}-\frac{{\rho}}{{B}}\int\frac{\rho_H}{\lambda Z^{(0)}} -\int\frac{B_HZ^{(0)}}{\lambda}\right)\left(\frac{1}{T}\mathcal{U}_{13}-\frac{{s}}{{B}} \int\frac{\rho_H}{\lambda Z^{(0)}}\right)\,,
\cr
\alpha^{xy}&=-\bar{\alpha}^{yx}=\frac{s}{{B}}\,,\cr
\alpha^{yx}&=-\bar{\alpha}^{xy}=\frac{4\pi}{X}\left(\mathcal{U}_{12}-\frac{{\rho}}{{B}}\int\frac{\rho_H}{\lambda Z^{(0)}}-\int\frac{B_HZ^{(0)}}{\lambda}\right)\,.\label{alph1d}
\end{align}
Finally the components of $\bar\kappa$ have the form:
\begin{align}
\bar{\kappa}^{xx}&=\frac{16\pi^2 T}{X}\,,\cr
\bar{\kappa}^{yy}&=\frac{1}{T}\mathcal{U}_{33}+\frac{{s}^2T}{{B}^2}\int\frac{\gamma^{1/2}}{\lambda^{1/2}Z^{(0)}}-\frac{T}{X}\left(\frac{1}{T}\mathcal{U}_{13}-\frac{{s}}{{B}}\int\frac{\rho_H}{\lambda Z^{(0)}}\right)^2\,,\cr
\bar{\kappa}^{xy}&=-\bar{\kappa}^{yx}=-\frac{4\pi T}{X}\left(\frac{1}{T}\mathcal{U}_{13}-\frac{{s}}{{B}} \int\frac{\rho_H}{\lambda Z^{(0)}}
\right)\,.\label{kap1d}
\end{align}

One can check that the following Onsager relations are satisfied:
$\alpha^T (B_H,\chi )=\bar\alpha(-B_H,-\chi )$, $\bar\kappa^T(B_H,\chi )=\bar\kappa(-B_H,-\chi )$
and $\sigma^T(B_H,\chi )=\sigma(-B_H,-\chi )$. 
Finally we note that
from these explicit expressions, one can easily obtain explicit expressions for $\sigma_{Q=0}$ and $\kappa$.
In this case one can explicitly check that $\sigma^{ij}_{Q=0}\ne \Sigma^{ij}_H$ as defined in \eqref{Hquants}.
For example $\sigma^{xy}_{Q=0}=(4\pi)^2T/(BXdet\bar\kappa)\int u_3(\rho s_H-s\rho_H)\ne 0$.

We can also calculate the thermal conductivity $\kappa$. As for the Q-lattice we find that the result can be obtained
by making the substitution $B_H\to -\rho_H$, $\rho_H\to B_H$ and $Z^{(0)}\to 1/Z^{(0)}$, the reason for which will be clarified in
appendix \ref{sduality}.

\section{Discussion}\label{sec7}
It was shown in \cite{Donos:2015gia,Banks:2015wha} (building on the earlier work in \cite{Donos:2014uba,Donos:2014cya,Donos:2014yya}) that the DC thermoelectric conductivity 
for a general class of static holographic lattice black hole solutions can be obtained by solving a system of
generalised Stokes equations on the black hole horizon\footnote{The formalism also allows for the possibility of multiple black hole
horizons.}. In this paper we have extended these
results from the static class to also include stationary black holes. A particularly interesting feature of this more general class
of black hole solutions is that they are dual to field theories that can have local electric and heat magnetisation 
currents. 

We showed that the resulting generalised Stokes equations on the black hole horizon contain some new terms that have a form that
could have been anticipated based on magneto hydrodynamics as well as some additional novel terms.
In deriving our results it was particularly important to identify the correct transport current fluxes after suitably
subtracting contributions arising from the magnetisation currents, in the spirit of
\cite{PhysRevB.55.2344} and generalising what was done for constant magnetic fields in \cite{Blake:2015ina,Hartnoll:2007ih}.

The results here and in \cite{Donos:2015gia,Banks:2015wha} 
provide a new avenue for systematically investigating DC conductivity and transport within holography.
For example, it may be possible to place bounds on the conductivity, extending
the results obtained for the electric conductivity in $D=4$ Einstein-Maxwell theory in \cite{Grozdanov:2015qia}.
It would also be interesting to further develop the relationship with studies of hydrodynamics as in, for example, 
\cite{Hartnoll:2007ih,Davison:2014lua,Davison:2015bea,Lucas:2015lna,Blake:2015hxa,Lucas:2015sya}. 
Making connections with hydrodynamical behaviour in graphene and other materials, as recently discussed in 
\cite{2015arXiv150800836L,2015arXiv150904165B,2015arXiv150905691M,2015arXiv150904713C}, is
another avenue for further investigation.

The Stokes equations that we have obtained have a natural extension to Navier-Stokes equations which contain time derivatives as
well as non-linear terms. It would be interesting if our formalism can be extended in such a way that these more general
equations can be used to extract information about the AC conductivities and possibly
conductivities beyond the level of linear-response (e.g. see \cite{Horowitz:2013mia}).

\section*{Acknowledgements}
We thank Elliot Banks, Mike Blake, Andrew Lucas and Subir Sachdev for helpful discussions.
The work is supported by STFC grant ST/J0003533/1, EPSRC grant EP/K034456/1,
and by the European Research Council under the European Union's Seventh Framework Programme (FP7/2007-2013), ERC Grant agreement ADG 339140. This research was supported in part by Perimeter Institute of Theoretical Physics. 
Research at Perimeter Institute is supported by the Government of Canada through Industry Canada and by the Province of Ontario through the Ministry of Economic Development \& Innovation.

\appendix

\section{The time-like KK reduction}\label{tlkkred}
We consider the metric and gauge-field
\begin{align}\label{gfanap}
ds^2&=-H^2 (dt+\alpha +t \zeta)^2+ds^2(M_{D-1})\,,\nn
A&=a_t(dt+\alpha+t\zeta)-tE +\beta\,,
\end{align}
with $H,\alpha,\beta,E,\zeta$ defined on $M_{D-1}$. We demand that
\begin{align}
d\zeta=0,\qquad dE=\zeta\wedge E\,.
\end{align}
In this appendix, we will not be working at linearised order in $E,\zeta$.

Locally we can eliminate all the time-dependence. To see this we write
$\zeta= dz$ and $E=e^z d e$, for locally defined functions $e,z$.
We now write $t=e^{-z}\tilde t$ so that $dt+t\zeta=e^{-z}d \tilde t$
and also do the gauge transformation $A\to A +d(\tilde t e)$.
Note that $\partial_t =e^z\partial_{\tilde t}$
with $\partial_{\tilde t}$ a Killing vector.

We now introduce the orthonormal frame
\begin{align}
e^0 =H(dt+\alpha+t\zeta)\,,\qquad
e^a=\bar e^a\,,
\end{align}
with $\bar e^a\bar e^a= ds^2(M_{D-1})$.
After some calculation
we find the following components of the Ricci tensor in this frame:
\begin{align}\label{tricci}
R_{00}&=\nabla^a(\nabla_a\ln H-\zeta _a)+(\nabla_a\ln H-\zeta _a)(\nabla^a\ln H-\zeta ^a)
+\frac{H^2}{4}(d\alpha-\alpha\wedge \zeta)^2\,,\nn
R_{b0}&=\frac{1}{2}\nabla_a \left(H(d\alpha-\alpha\wedge\zeta)^a{}_b\right)
+\left(\nabla_a \ln H-\zeta_a\right)H(d\alpha-\alpha\wedge\zeta)^a{}_b\,,\nn
R_{ab}&=\bar R_{ab}+\frac{H^2}{2}(d\alpha-\alpha\wedge\zeta)^2_{ab}
-\nabla_a(\nabla_b\ln H-\zeta_b)
-(\nabla_a\ln H -\zeta_a)(\nabla_b\ln H -\zeta_b)\,.
\end{align}
The Ricci scalar can then be written
\begin{align}\label{expRSap}
R=\bar R+\frac{H^2}{4}(d\alpha-\alpha\wedge \zeta)^2-
2\nabla^a(\nabla_a\ln H-\zeta _a)-2(\nabla_a\ln H-\zeta _a)(\nabla^a\ln H-\zeta ^a)\,.
\end{align}
We also have
\begin{align}\label{fexp}
F&=\frac{1}{H}(da_t-a_t\zeta+E)\wedge e^0+a_td\alpha+d\beta+\alpha\wedge(E-a_t\zeta)\,,
\end{align}
For an arbitrary two-form $T=T_1\wedge e^0+T_2$, built from $T_1$ and $T_2$ on $M_{D-1}$, we have
$*T=(-1)^D\bar * T_1+e^0\wedge \bar * T_2$
where $\bar *$ is the Hodge dual with respect to the metric $ds^2(M_{D-1})$.
Thus we have
\begin{align}\label{stareff}
*F=(-1)^D\bar *\frac{1}{H}(da_t-a_t\zeta+E)+e^0\wedge \bar * \left[
a_td\alpha+d\beta+\alpha\wedge(E-a_t\zeta)\right]\,.
\end{align}
Similar expressions can be written down for $G$ and $*G$.

\section{Holographic renormalisation}\label{holren}
In this section we will show that the counterterms do not make any contribution to
the transport current flux densities. 
For simplicity we will restrict our detailed considerations
to the cases of $D=4$ and $D=5$ bulk spacetime dimensions and just to Einstein-Maxwell theory.
We will also restrict to the configurations discussed in section \ref{sec4} to illustrate the main idea.

For the two cases, we need to consider the following actions (e.g. \cite{Henningson:1998gx,Balasubramanian:1999re,Taylor:2000xw})
\begin{align}
S_{D=4}&=S_{\mathrm{bulk}}+S_{\mathrm{GH}}+S_{\mathrm{count}}\,,\nn
S_{D=5}&=S_{\mathrm{bulk}}+S_{\mathrm{GH}}+S_{\mathrm{count}}+S_{\mathrm{log}}\,,
\end{align}
where $S_{\mathrm{bulk}}$ is the bulk action given in \eqref{eq:bulk_action} (with $\phi=0$) 
and $S_{\mathrm{GH}}$ the Gibbons-Hawking contribution. The counterterm for both cases is given by
\begin{align}
S_{\mathrm{count}}=-\int\,d^{D-1}x\,\sqrt{-\gamma}\,\left(2(D-2)+\frac{1}{D-3}R_{D-1}\right)\,,
\end{align}
where $\gamma_{\mu\nu}$ is the metric on the hypersurfaces normal to the outward pointing $n^{\mu}$ on the conformal boundary and $R_{D-1}$ is the corresponding Ricci scalar. In $D=5$ we also have a logarithmic term
\begin{align}\label{eq:log_term}
S_{\mathrm{log}}=-\frac{1}{4}\,\ln r\,\int d^4x\,\sqrt{-\gamma}\,\left(  \left(R_{4}\right)_{\mu\nu}\left(R_{4}\right)^{\mu\nu}-\frac{1}{3}\,R_{4}^{2}-F^{2}\right)\,,
\end{align}
where we have implicitly assumed a Fefferman-Graham expansion close to the boundary at $r\to\infty$.
For the metric and gauge-field on the boundary, we write:
\begin{align}\label{eq:b_metric}
ds_{D-1}^{2}&=-\tilde{H}^2\,\left(dt+\tilde{\alpha}+t\,\zeta\right)^{2}+\tilde{h}_{ij}dx^{i}dx^{j}\,,\nn
A_{D-1}&=\tilde a_t\left(dt+\tilde{\alpha}+t\,\zeta\right)-t\,E+\tilde{\beta}\,.
\end{align}

Let us first consider the contributions of $S_{count}$ to the holographic currents. Clearly this will
not give any contribution to the renormalisation of the current of the background
$J^{(B)i}_\infty$ but it will to $Q^{(B)i}_\infty$. However,
we will see that this can be absorbed by renormalising $M^{ij}_T$ in \eqref{caljdefs}, rendering it finite.
Recalling that $Q^i 
=-2{\pi^i}_t-\pi^iA_t$ we want to find the counterterm contribution to\footnote{In our conventions the stress tensor of the dual field theory
is obtained as the on-shell variation of the action: $t^{\mu\nu}=(2/\sqrt{-\gamma})\delta S/\delta \gamma_{\mu\nu}$. We also have
sign change $t^{\mu\nu}= 2(\pi^{\mu\nu})/\sqrt{-\gamma}$}
$-2\pi^i{}_t$. There will be no contributions
from terms involving $\gamma^i{}_{t}$, which of course vanish, and we find
\begin{align}
\left(\pi^{i}{}_{t}\right)_{\mathrm{count}}&=\frac{\sqrt{-\gamma}}{(D-3)}\, \left(R_{D-1}\right)^{i}{}_{t}\,.
\end{align}
For the metric \eqref{eq:b_metric}, using \eqref{tricci} we have
\begin{align}
\sqrt{-{\gamma}}\,\left(R_{D-1}\right)^{i}{}_{t}&=-\frac{1}{2}\tilde{h}^{1/2}\,\tilde{\nabla}_{j}\left(\tilde{H}^{3}\,\left( d\tilde\alpha-\tilde{\alpha}\wedge \zeta\right)^{ij} \right)+\tilde{h}^{1/2}\,\tilde{H}^{3}\,\left( d\tilde\alpha-\tilde{\alpha}\wedge \zeta\right)^{ij}\,\zeta_{j}
\end{align}
where on the right hand side indices have been raised with $\tilde h^{ij}$ and we have kept non-linear terms in $\zeta$. 
Thus the counterterm contribution to the heat current can be written as
\begin{align}
Q^i_{count}=\partial_j\delta M^{ij}_T-2\delta M^{ij}_T\zeta_j\,,
\end{align}
where $\delta M^{ij}_T=\tilde{h}^{1/2}\,\tilde{H}^{3}\,\left( d\tilde\alpha-\tilde{\alpha}\wedge \zeta\right)^{ij}/(D-3)$.
One can now check that if we add $\delta M^{ij}_T$ to $M^{ij}_T$ in \eqref{caljdefs} then it precisely cancels
the divergence proportional to $r^{D-3}$ at $r\to\infty$. Similar comments apply to the contributions of the log terms in $D=5$, which renormalise both the electric
and the heat currents.
Thus, in effect, we can replace the right hand side
of \eqref{caljdefs} with the finite renormalised quantities, leaving the left hand side (which is our principle interest)
unchanged.

\section{Magnetisation comments}\label{magcom}
We make some additional comments concerning the relationship between
$m$ and $m_T$, defined in \eqref{defemms}, and magnetisation and heat magnetisation of the equilibrium black holes, respectively. 
To do this we rewrite the bulk action \eqref{eq:bulk_action} for the background metric using the KK decomposition on the time direction.
We are interested in the terms in the action that involve the field strengths $d\alpha$ and $d\beta$ (where we are now using the same notation for the background solution as we did for the perturbed solution in the previous appendix). After some calculation we find
\begin{align}
S=\int dt\wedge\left[\frac{H^3}{2}\bar * d\alpha\wedge d\alpha-\frac{ZH}{2}\bar * (a_t d\alpha+d\beta)\wedge(a_t d\alpha+d\beta)+\dots \right]\,.
\end{align}
To obtain this we used the expression \eqref{expRSap} for the Ricci scalar (upon setting $\zeta=0$).
We Euclideanise via $t=-i\tau$ and $I=-iS$. After recalling that the bulk contribution
to the free energy is given by $W_{bulk}=TI$ (on-shell) we get
\begin{align}
W_{bulk}=-\int_{M_{D-1}} \left[\frac{H^3}{2}\bar * d\alpha\wedge d\alpha-\frac{ZH}{2}\bar * (a_t d\alpha+d\beta)\wedge(a_t d\alpha+d\beta)+\dots \right]\,.
\end{align}

We next note that the closed forms $d\alpha$ and $d\beta$ on $M_{D-1}$ have exact pieces as well as some harmonic pieces
(since $\alpha$ and $\beta$ do not have to be globally defined). We introduce a basis of harmonic two-forms, $\omega^A$, on $M_{D-1}$
that approach harmonic two-forms, $\omega^{A}_{\infty}$, on $\Sigma_d$ at the AdS boundary. We can then write
$d\alpha=\sum_Aa_A\omega^A+\dots$ and $d\beta= \sum_A b_A\omega^A+\dots$, where $a_A,b_A$ are constants and the dots refer to 
the exact pieces.
We can then define the magnetisation, $\bar M_{\omega^A}$, and the thermal magnetisation, $\bar M_{T\omega^A}$, as the variation of the on-shell action with respect $a^A,b^A$ finding the following integrals over the full bulk space (at constant $t$): 
\begin{align}\label{defM}
\bar M_{\omega^A}\equiv\frac{\delta W}{\delta b_A}&=-\int_{M_{D-1}}m\wedge \omega^A\,,\nn
\bar M_{T\omega^A}\equiv\frac{\delta W}{\delta (-a_A)}&=-\int_{M_{D-1}} m_T\wedge \omega^A\,,
\end{align}
where $m$ and $m_T$ were defined in \eqref{defemms}. Recalling that $dm=dm_T=0$ on shell, and that $m$ and $m_T$
vanish at black hole horizons, we see that the constants $\bar M_{\omega^A}$, $\bar M_{T\omega^A}$, which 
are properties of the equilibrium CFT, do depend on the choice of harmonic form $\omega^A_\infty$ 
on $\Sigma_d$ of the AdS boundary. Indeed if we shift $\omega^A\to\omega^A+d\lambda$ we see that the result
will depend on $\lambda_\infty$.

We can now make a connection with the formulae for the transport current flux densities given in \eqref{pdexps}. 
Indeed we can write
\begin{align}\label{pdexps2}
\bar{\mathcal{J}}^a
&=-\int_{\Sigma_d}\eta_\infty^a\wedge i_{k}*Z(\phi)F  -\bar M_{\eta^a\wedge\zeta}\,,\,,\nn
\bar{\mathcal{Q}}^a&= -\int_{\Sigma_d}\eta_\infty^a\wedge i_{k}*G  -\bar M_{\eta^a\wedge E} -2\bar M_{T\eta^a\wedge\zeta}\,,
\end{align}
and we recall that the left hand side is independent of the representative chosen for the harmonic one-form $\eta^a_\infty$.

\section{The one-dimensional lattice calculation}\label{onedapp}
Here we present a few details how to obtain the conductivities for the one-dimensional lattice
that we presented in section \eqref{onedtext}.
We start by averaging \eqref{2keyeq} over a period in the $x$ direction. We immediately obtain
\begin{align}\label{2jexps}
{J^x}_{(0)}=\frac{{\rho}}{{B}}E_y
+\frac{{sT}}{{B}} \zeta_y\,,
\end{align}
where $\rho$, $s$, $B$ were defined in \eqref{rsb}.
Integrating \eqref{2keyeq} gives:
\begin{align}\label{2firstintegral}
\partial_x v^y
=\frac{\gamma^{1/2}}{\lambda^{3/2}}(\chi  v_0+w_1E_y+w_2\zeta_y+c_1)\,,
\end{align}
where $c_1$ is a constant of integration and the periodic functions $w_i$ are given in \eqref{defws}.
Averaging \eqref{2firstintegral} over a period, we obtain the value of $c_1$:
\begin{align}
c_1\int\frac{\gamma^{1/2}}{\lambda^{3/2}}=-\int\frac{\gamma^{1/2}}{\lambda^{3/2}}(\chi  v_0-w_1E_y-w_2\zeta_y)\,.
\end{align}
Integrating \eqref{2firstintegral} gives:
\begin{align}\label{2secondintegral}
 v^y
&=\int^x\frac{\gamma^{1/2}}{\lambda^{3/2}}(\chi  v_0+w_1E_y+w_2\zeta_y+c_1)+c_2\,,\cr
&=u_1 v_0+u_2E_y+u_3\zeta_y+c_2\,,
\end{align}
where the periodic functions $u_i$ are given in \eqref{defus}
and $c_2$ is a constant of integration. Observe that the $u_i$ satisfy:
\begin{align}
\partial_x(\frac{\lambda^{3/2}}{\gamma^{1/2}}\partial_xu_1)&=\partial_x\chi \,,\cr
\partial_x(\frac{\lambda^{3/2}}{\gamma^{1/2}}\partial_xu_2)&=\partial_xw_1=\frac{{\rho}}{B}{B_H}-\rho_H\,,\cr
\partial_x(\frac{\lambda^{3/2}}{\gamma^{1/2}}\partial_xu_3)&=\partial_xw_2=\frac{Ts}{B}B_H-Ts_H\,.\label{uidentities}
\end{align}

Then, using \eqref{2jexps} and \eqref{2secondintegral}, we can rewrite \eqref{2jxeqn} as:
\begin{align}
\frac{\gamma^{1/2}}{\lambda^{1/2}Z^{(0)}}(\frac{{\rho}}{{B}}E_y
+\frac{{sT}}{{B}}\zeta_y)&=&E_x+\partial_xw+\frac{\rho_H}{\lambda Z^{(0)}}v_{0}+B_H(u_1 v_0+u_2E_y+u_3\zeta_y+c_2)\,.
\end{align}
Averaging this over a period, we can determine the value of $c_2$
\begin{align}
c_2&=&(\frac{{\rho}}{{B}^2}\int\frac{\gamma^{1/2}}{\lambda^{1/2}Z^{(0)}}-\frac{1}{{B}}\int B_H u_2)E_y+(\frac{{s}T}{{B}^2}\int\frac{\gamma^{1/2}}{\lambda^{1/2}Z^{(0)}}- \frac{1}{{B}}\int B_H u_3)\zeta_y\cr
&&\qquad\qquad\qquad-\frac{E_x}{{B}}-(\frac{1}{{B}}\int\frac{\rho_H}{\lambda Z^{(0)}}+\frac{1}{{B}}\int B_H u_1 )v_0\,.
\end{align}
Substituting this back into \eqref{2secondintegral} gives:
\begin{align}\label{2vyexp}
 v^y&=\left(u_2+\frac{{\rho}}{{B}^2}\int\frac{\gamma^{1/2}}{\lambda^{1/2}Z^{(0)}}-\frac{1}{{B}}\int B_Hu_2\right)E_y+\left(u_3+\frac{\bar{s}T}{{B}^2}\int\frac{\gamma^{1/2}}{\lambda^{1/2}Z^{(0)}}- \frac{1}{{B}}\int B_H u_3\right)\zeta_y\nn
&\qquad\qquad\qquad-\frac{E_x}{{B}}+\left(u_1-\frac{1}{{B}}\int\frac{\rho_H}{\lambda Z^{(0)}}-\frac{1}{{B}}\int B_H u_1 \right)v_0\,.
\end{align}
Substituting this into \eqref{2jyeqn} gives:
\begin{align}
{J^y}^{(0)}&=\left(\frac{\gamma^{1/2}Z^{(0)}}{\lambda^{1/2}}+\rho(u_2+\frac{{\rho}}{{B}^2}\int\frac{\gamma^{1/2}}{\lambda^{1/2}Z^{(0)}}-\frac{1}{{B}}\int B_Hu_2)\right)E_y\cr
&\qquad\qquad\qquad+\rho\left(u_3+\frac{\bar{s}T}{{B}^2}\int\frac{\gamma^{1/2}}{\lambda^{1/2}Z^{(0)}}- \frac{1}{{B}}\int B_H u_3\right)\zeta_y-\frac{\rho}{{B}}E_x\cr
&\qquad\qquad\qquad+\left(\rho(u_1-\frac{1}{{B}}\int\frac{\rho_H}{\lambda Z^{(0)}}-\frac{1}{{B}}\int B_Hu_1 )-\frac{BZ^{(0)}}{\lambda}\right)v_{0}\,.
\end{align}
Finally, averaging \eqref{2navx} over a period in the $x$ direction and using \eqref{2jyeqn}, \eqref{2jexps} and \eqref{2secondintegral}, we obtain an expression for the constant $v_0$:
\begin{align}
v_{0}&=\frac{1}{X}\left(\frac{{\rho}}{{B}}
\int\frac{\rho_H}{\lambda Z^{(0)}}+\int u_2\partial_x\chi +\int\frac{B_HZ^{(0)}}{\lambda}\right)E_y\cr
&\qquad\qquad\qquad+\frac{1}{X}\left(\frac{{s}}{{B}}T \int\frac{\rho_H}{\lambda Z^{(0)}}+\int u_3 \partial_x\chi \right)\zeta_y+\frac{4\pi T}{X}\zeta_{x}\,,
\end{align}
where the constant $X$ is given by
\begin{align}
X&=\int Y+\int\frac{B^2_HZ^{(0)}}{\gamma^{1/2}\lambda^{3/2}}-\int u_1\partial_x\chi \,.
\end{align}

Therefore we have successfully solved for the four quantities $v_0$, ${J^x}^{(0)}$, $v^y$ and ${J^y}^{(0)}$ in terms of the sources. 
Since ${Q^i}_{(0)}=Ts_Hv^i$, we have ${\bar{Q}^x}_{(0)}=Q^x_{(0)}=4\pi T v_0$ and ${\bar{Q}^y}{}^{(0)}=4\pi T\int(\gamma\lambda)^{1/2}v^y$ while
${\bar{J}^x}_{(0)}=J^x_{(0)}$ and ${\bar{J}^y}_{(0)}=\int J^y_{(0)}$. After introducing the matrix 
\begin{align}
\mathcal{U}_{ij}\equiv\int\frac{\lambda^{3/2}}{\gamma^{1/2}}\partial_xu_i\partial_xu_j\,,
\end{align}
and a little effort, one can obtain the expressions given in \eqref{sig1d}-\eqref{kap1d}. Observe that using \eqref{uidentities} we can simplify the
expressions for $\mathcal{U}_{ij}$ a little. We also note that under the transformation
$(B_H,\chi )\to (-B_H,-\chi )$ we have $u_1$ is odd and $u_2,u_3$ are even, which is useful for checking the Onsager relations.

\section{S-duality transformations of conductivity}\label{sduality}
For the special case of holographic models in $D=4$ there is an
important S-duality invariance of the generalised Stokes equations \eqref{auxstokes}. We emphasise that
this independent of whether the bulk theory itself exhibits S-duality.

For the background black holes we write $B_H=F^{(0)}_{xy}$ and 
$\rho_H=\sqrt{h^{(0)}}Z^{(0)}A_t^{(0)}$, as before. Consider transforming
the background quantities via
\begin{align}\label{hortrans}
B_H\rightarrow\rho_H\,,\qquad
\rho_H\rightarrow-B_H\,,\qquad
Z^{(0)}\rightarrow 1/Z^{(0)}\,,
\end{align}
and also transforming the perturbation via
\begin{align}
E_i+\partial_iw\rightarrow-\epsilon(ij)J^{j}_{(0)}\,,\qquad
\epsilon(ij)J^{j}_{(0)}\rightarrow E_i+\partial_iw\,,
\label{transformationeq2}
\end{align}
where $\epsilon(xy)=1$.
Note that $\rho_H(E_i+\partial_iw)+\epsilon(ij)J^j_{(0)}B_H$ is left invariant under these transformations, as is
$\partial_i J^i_{(0)}$, and hence the Stokes equations \eqref{auxstokes} are also.

To deduce how the conductivities change under this transformation we now consider the case that $(x,y)$ parametrise a torus.
We first observe that $\partial_iw$ in \eqref{transformationeq2} make no contribution to the current flux densities
defined in \eqref{avecur}. Thus we deduce 
$\bar{E}_i\rightarrow-\epsilon(ij)\bar{J}^{j}_{(0)}$ and $\bar{J}^{i}_{(0)}\rightarrow-\epsilon(ij)\bar{E}_j$.
Next we examine how the definitions $\bar{J}^i_{(0)}=\sigma^{ij}\bar{E}_j+T\alpha^{ij}\bar{\zeta}_j$
and $\bar{Q}^i_{(0)}=T\bar{\alpha}^{ij}\bar{E}_j+T\bar{\kappa}^{ij}\bar{\zeta}_j$ transform. After a little rearrangement
we deduce that the conductivities transform as
\begin{align}\label{dualtc}
\sigma^{ij}&\rightarrow-\epsilon(ik)\sigma^{-1}_{kl}\epsilon(lj)\,,\qquad
\alpha^{ij}\rightarrow-\epsilon(ik)\sigma^{-1}_{kl}\alpha^{lj}\,,\cr
\bar{\alpha}^{ij}&\rightarrow-\bar{\alpha}^{ik}\sigma^{-1}_{kl}\epsilon(lj)\,,\qquad
\bar{\kappa}^{ij}\rightarrow \kappa^{ij}\equiv\bar{\kappa}^{ij}-T\bar{\alpha}^{ik}\sigma^{-1}_{kl}\alpha^{lj}\,.
\end{align}
Notice that the transformation of $\alpha$ involves the Nernst response, $\vartheta\equiv -\sigma^{-1}\alpha$, which determines
the electric field that is generated by a heat gradient at zero current flow via $E_{J=0}=T\vartheta \zeta$.

It is important to realise that this transformation is relating conductivities as functions of local horizon data
$(\rho_H,B_H,Z^{(0)})$ to conductivities with transformed local horizon data $(-B_H,\rho_H,1/Z^{(0)})$. In general these
are not the same as the holographic physical quantities defined at the AdS boundary. Furthermore, one is not guaranteed that
there is in fact a black hole solution that has the transformed horizon quantities. It is interesting to
observe, however, that if we want to obtain $\kappa$ for a horizon with $(\rho_H,B_H,Z_0)$ we simply need to
take the expression for $\bar\kappa$ for a horizon with $(\rho_H,B_H,Z^{(0)})$ and then apply the inverse 
transformation directly $B_H\rightarrow-\rho_H$, 
$\rho_H\rightarrow B_H$ and $Z^{(0)}\rightarrow 1/Z^{(0)}$.
This explains the observations concerning $\kappa$
made at the end of section \ref{comparsec} and \ref{onedtext},
for the case of the Q-lattice and also the one-dimensional lattice, respectively.
Similarly, for the Nernst response we have $\vartheta(\rho_H,B_H,Z^{(0)})=-\epsilon\alpha(B_H,-\rho_H,1/Z^{(0)})$.

Now consider the special class of $D=4$ models which have equations of motion that are
invariant under the S-duality transformations
\begin{align}
{F}_{\mu\nu}&\rightarrow Z(\phi)*{F}_{\mu\nu}\,,\cr
\phi&\rightarrow-\phi\,,
\end{align}
with the metric left invariant. 
This requires $V$ to be an even function of $\phi$ and
$Z(-\phi)=Z(\phi)^{-1}$ which is achieved if $Z=e^f$, with $f$ an odd function of $\phi$.
By examining these duality transformation at the horizon
and using $\epsilon_{trxy}=\sqrt{-g}$ we see that induces \eqref{hortrans} on the background.
In other words, the bulk S-duality transformation is mapping one bulk black hole solution with
$(\rho_H,B_H,Z_0)$ to another solution with $(-B_H,\rho_H,1/Z_0)$.

After performing the S-duality transformation \eqref{dualtc} twice, we see that the conductivities transform as
\begin{align}\label{dualtwice}
\sigma^{ij}&\rightarrow\sigma^{ij}\,,\qquad
\alpha^{ij}\rightarrow-\alpha^{ij}\cr
\bar{\alpha}^{ij}&\rightarrow-\bar{\alpha}^{ij}\,,\qquad
\bar{\kappa}^{ij}\rightarrow\bar{\kappa}^{ij}\,.
\end{align}
This transformation arises from the symmetry transformation $A_\mu \rightarrow -A_\mu$ of the full bulk equations of motion
which is also valid in any spacetime dimension and for any choice of $Z,V$.

Thus, for general theories in arbitrary spacetime dimension if we have solutions with $\rho_H=0$, $B_H=0$, we can use
\eqref{dualtwice} to conclude that we must have $\alpha^{ij}=\bar{\alpha}^{ij}=0$. For general theories in $D=4$ spacetime dimensions, if we
have solutions with $\rho_H=0$, $B_H=0$ and $Z^{(0)}=1$ we can use \eqref{dualtc} to conclude that in
addition we have $\det(\sigma^{ij})=1$.

We note that the transformation for $\sigma$ given in \eqref{dualtc}
appeared in \cite{Grozdanov:2015qia} for the case of pure $D=4$ Einstein-Maxwell theory, 
which has bulk S-duality. In addition it was shown for this specific model in \cite{Grozdanov:2015qia}
that solutions with $\rho_H=0$, $B_H=0$ have $\det(\sigma^{ij})=1$.

\section{Analytic Black Hole solutions with NUT charges}\label{nutchge}

We consider $D=4$ black hole solutions given by the ansatz:
\begin{align}\label{nutty}
ds^2&=-U(r)\left[dt+k(xdy-y dx)\right]^2+\frac{dr^2}{U(r)}+e^{2V(r)}(dx^2+dy^2)
\end{align}
with $A_\mu=\phi=0$ and we set in $V=-6$ in \eqref{eq:bulk_action}. 
It is straightforward to solve the equations of motion for $U,V$ and we find
\begin{align}
U&=\frac{ r^4+ 6k^2 r^2 -3k^4  -2m r}{k^2 +  r^2}\,,\nn
e^{2V}&=(k^2+r^2)\,.
\end{align} 
where $m$ is a constant. 
The solutions approach $AdS_4$ as $r\to\infty$ with 
$d\chi^{(\infty)}=-2k dx\wedge dy$, in the notation of \eqref{asmet},
corresponding to a non-trivial constant NUT charge. Note that a holographic analysis 
of Euclidean AdS Taub-NUT solutions with spherical spatial sections, rather than
the flat spatial sections of \eqref{nutty}, was carried out in \cite{Chamblin:1998pz}.

An important point to notice about these solutions is that they have closed time-like curves everywhere in the spacetime including
the boundary. To see this we switch from cartesian coordinates $(x,y)$ to polar coordinates $(\rho,\theta)$ and then notice that
the Killing vector $\partial_\theta$, which has closed orbits with period $2\pi$, becomes timelike on any constant $r$ hypersurface
for $\rho^2\ge \frac{e^{2V}}{k^2 U}$.

We choose $m\ge 0$. There is a black hole horizon located at the largest root of $U(r)$. Shifting $r\to r+r_0$ we can choose
the horizon to be located at $r=0$ by choosing the appropriate root of the quartic: $r_0^4+6k^2r_0^2-3k^4-2mr_0=0$, with $r_0>0$.
The temperature is given by $T=(3/4\pi r_0)(k^2+r_0^2)$ and the entropy density is $s=(4\pi)^2Tr_0/3$. 
The Stokes equations, which have $d\chi^{(0)}=-(4\pi T)2k dx\wedge dy$, can be solved in
a similar manner to the Q-lattices in section \ref{rhotot}, and we find
\begin{align}\label{chiDCcond}
\sigma^{ij}=Z^{(0)}\left(\begin{array}{cc} 1 & 0 \\ 0 &1\end{array}\right)\,,\qquad
\bar{\kappa}^{ij}=\,\frac{s}{2k} \left(\begin{array}{cc}0&  -1\\ 1 & 0\end{array}\right)\,,
\end{align}
and $\alpha^{ij}= \bar\alpha^{ij}= 0$.


\providecommand{\href}[2]{#2}\begingroup\raggedright\endgroup

\end{document}